\documentclass[11pt,a4paper]{iopart}

\usepackage{cite}
\usepackage{graphicx} 
\usepackage{xcolor}

\usepackage{braket}

\usepackage[utf8]{inputenc}
\usepackage{subcaption}

\usepackage{iopams}

\expandafter\let\csname equation*\endcsname\relax

\expandafter\let\csname endequation*\endcsname\relax 

\usepackage{amsmath}
\usepackage{amssymb}
\usepackage{hyperref}

\usepackage{fancyhdr}
\makeatletter
\def\@mkboth#1#2{}
\newlength\appendixwidth

\preto\appendix{\addtocontents{toc}{\protect\patchl@section@andsubsection}}

\newcommand{\patchl@section@andsubsection}{%
  \settowidth{\appendixwidth}{\textbf{Appendix }}%
  \addtolength{\appendixwidth}{1.5em}%
  \patchcmd{\l@section}{1.5em}{\appendixwidth}{}{\ddt}%
  \patchcmd{\l@subsection}{2.3em}{\dimexpr\appendixwidth+1.8em\relax}{}{\ddt}%
}
\makeatother

\begin{document}

\title{Weakly interacting Bose-Einstein condensate with stochastic resetting}

\author{Nikhil Mesquita}
\address{Raman Research Institute, Bangalore 560080, India\\
Email: nikhilm@rrimail.rri.res.in}

\author{Manas Kulkarni}
\address{International Centre for Theoretical Sciences, Tata Institute of Fundamental Research,
Bangalore 560089, India\\
Email: manas.kulkarni@icts.res.in }

\author{Satya N. Majumdar}
\address{LPTMS, CNRS, Universit\'e Paris-Sud, Universit\'e Paris-Saclay, 91405 Orsay, France\\
Email: satya.majumdar@universite-paris-saclay.fr}

\author{Sanjib Sabhapandit~\footnote{Author to whom any correspondence should be addressed.}}
\address{Raman Research Institute, Bangalore 560080, India\\
Email: sanjib@rri.res.in}

\date{\today}

\begin{abstract}
Stochastic resetting can generate strong correlations in many-body systems through a shared fluctuating environment. Here, we investigate how such dynamically emergent correlations (DEC) coexist with intrinsic (direct) interactions in a weakly repulsive Bose-Einstein condensate described within the Gross-Pitaevskii framework. The condensate undergoes free expansion from an initial Thomas-Fermi state and is stochastically reset to this initial state at a constant rate. We show that the resetting protocol drives the system into a unique nonequilibrium steady state and obtain its density profile, edge statistics, and full counting statistics analytically. The steady-state density retains an inverted-parabolic form within the core, while developing exponentially decaying tails outside it. We further show that resetting induces nontrivial fluctuations of the condensate edge and particle number, yielding exact scaling forms for the edge distribution and the full counting statistics. Our results provide a tractable setting for exploring the interplay between intrinsic repulsive interactions and attractive DEC generated by a common stochastic environment in quantum many-body systems. 
\end{abstract}

\maketitle

\newpage
\noindent\rule{\hsize}{2pt}
\tableofcontents
\noindent\rule{\hsize}{2pt}

\section{Introduction}
\label{s:intro}
Recent years have seen a growing focus on the study of noninteracting many-particle systems that are influenced by a fluctuating shared environment~\cite{BLM2023,BLM2024, BLM2024_2, SM2024, MMS2025, dBM2026, DS2025, BMS2025, Olsen2026, BCK2025,   deMauro_2026, galla2026, demauro2026effects, MCP2022, KM2023, KMS2025, DIG2025, MKM2025, MS2026}. The fluctuations in the environment dynamically generate strong correlations in the system, even though there are no  inherent direct interactions between particles. Despite these strong correlations, the joint probability density function (JPDF) of the concerned degrees of freedom $\{x_1, x_2, \dotsc, x_N\}$ assumes the so-called conditionally independent and identically distributed (CIID) form~\cite{MS2026, BLM2023} 
\begin{equation}
\label{eq:CIID}
    P(x_1, x_2, \ldots x_N) = \int _{-\infty}^{\infty} \, dV \,  h(V) \, \prod_{j=1}^{N} p(x_j|V) \, .
\end{equation}
Here, conditioned on a random variable $V$, the product indicates an ideal noninteracting gas where each degree of freedom is distributed by $p(x_j|V)$ with $V$ being a parameter. Finally, the JPDF of the full gas is obtained by averaging over the random variable $V$ drawn from its PDF $h(V)$. The random variable $V$ and its PDF $h(V)$ as well as the PDF $p(x_j|V)$ are, of course, model dependent. Moreover, the stochastic variable $V$ arises from the underlying fluctuating environment, which is independent of $\{x_i\}$. Of course, not every JPDF $P(x_1, x_2, \ldots x_N)$ can be expressed in a CIID form. Even when the JPDF can be expressed in a CIID form, it is not always easy to find, (i)  how the random variable $V$ conditions the PDF $p(x|V)$, and (ii) the PDF $h(V)$. In fact, in many problems this CIID structure is actually rather hidden and is not immediately manifest. Recently, this CIID structure has been found in the nonequilibrium stationary state (NESS) in  a number of many-body systems undergoing stochastic resetting dynamics or its variants.
The fact that the JPDF in Eq.~\eqref{eq:CIID} is not factorizable indicates that the particles are correlated (usually all-to-all attractive). Usually, computing any observable for a strongly correlated many-body state is hard, but the special CIID structure
in Eq.~\eqref{eq:CIID} makes it relatively straightforward: one first computes it for an ideal noninteracting gas with
the parameter V fixed (this is often very easy) and then one needs to
average this observable over $V$ drawn from $h(V)$. This procedure has
been exploited recently in a number of noninteracting models undergoing stochastic dynamics in both  classical~\cite{BLM2023,BLM2024,BLM2024_2, SM2024, MMS2025, dBM2026, DS2025, BMS2025, Olsen2026, BCK2025,   deMauro_2026, galla2026, demauro2026effects} and quantum~\cite{MCP2022, KMS2025, DIG2025, MKM2025} context.
For a recent perspective on the subject, see Ref.~\cite{MS2026}.

Most of the recent studies mentioned above  deal with noninteracting systems subjected to a common shared fluctuating environment such as simultaneous stochastic resetting. This shared environmental fluctuation makes the system strongly correlated in the stationary state even in the absence of interactions. Such correlations generated by dynamics have been coined ``dynamically emergent correlations" (DEC). However, in an interacting system, there are direct correlations between the particles which do not have a dynamical origin. Hence, when such an interacting system is subject to the shared environmental fluctuation, the nonequilibrium steady state (NESS) will have correlations between particles coming from both (i) from the shared environmental fluctuation and (ii) direct interaction. Therefore, it is natural to explore the interplay between the two sources of correlation in the NESS of such interacting many body systems.

Indeed, in a recent experiment,  colloidal beads in water trapped harmonically by optical lasers was studied in Ref.~\cite{BCK2025}, where the frequency of the trap was modulated telegraphically between two values $\mu_1$ and $\mu_2$ with constant rates. It was found that despite the presence of hydrodynamic interactions between the beads, the correlations between the particles in the stationary state as well as some other observables were matched very well with the predictions of the noninteracting theory~\cite{BLM2024_2}. This naturally raised the question under what conditions the DEC would dominate over the correlations that are generated by the direct interactions between the particles.

In most of the examples cited above, the environmental fluctuations, such as the simultaneous resetting or fluctuating the parameters of a trap  stochastically, create an effective \emph{attractive} correlations between particles. Hence, in these examples, it would be interesting to see what happens if there are inherently repulsive interaction between particles which may possibly counterbalance the DEC and this competition may lead to interesting NESS. So far such a NESS with competing DEC and repulsive interactions has been explored only in one example, namely, the Dyson Brownian gas in one dimension subject to simultaneous stochastic resetting~\cite{BMS2025_2}. Here, between two successive resettings the positions of the particles evolve via overdamped Brownian dynamics (classical) in the presence of a pairwise logarithmic repulsion between the particles. 
In this example, the interaction between any pair of particles is repulsive and long-ranged. 
Moreover the dynamics is classical.
However, in many realistic systems the repulsion is short-ranged, and moreover, the dynamics is governed by quantum unitary
evolution and not via classical dynamics.
Hence, it is interesting to explore the competition between short-ranged repulsive interactions and attractive DEC in 
quantum systems whose natural dynamics
is unitary.

In this paper, we consider a dilute three-dimensional gas of bosons with short-range
repulsive interactions, confined in an elongated (cigar-shaped) harmonic trap. For a large
number of weakly interacting particles, it is known~\cite{Dunjko_2, LS2001} that when
(i) the trap is highly elongated, (ii) its transverse width is much larger than the
$s$-wave scattering length, and (iii) the mean interparticle distance along the longitudinal direction 
is also much larger than the scattering length, the ground state of the three-dimensional
system takes a simple effective form: in the transverse directions, the gas is frozen in
the ground state of the transverse harmonic oscillator, while its longitudinal density
profile is given by the ground state of the one-dimensional Gross-Pitaevskii equation
(GPE), also known as the nonlinear Schr\"odinger equation
(NLS)~\cite{PS2016, Pitaevskii1961, Gross1963, LS2001, ELS2007}. In this regime, the gas is
dilute in three dimensions but dense along the longitudinal direction. We further consider the limit
in which this longitudinal ground state takes the Thomas-Fermi form.

Our protocol is as follows. Starting from this Thomas-Fermi state, we switch off the
longitudinal trap, keeping the transverse confinement on, and let the gas evolve for a
random time $\tau_1$ drawn from the exponential distribution $p(\tau)=r e^{-r\tau}$. At the end of this deterministic evolution, we reset the system to its initial state, 
the longitudinal trap is switched back on and the gas is instantaneously cooled to the
Thomas-Fermi state. We then draw a new interval $\tau_2$, independently from the same
distribution, let the gas expand again for a time $\tau_2$, and reset it once more to the Thomas-Fermi initial state. This cycle of alternating deterministic evolution up to a random time $\tau$ followed by the instantaneous resetting to the Thomas-Fermi initial state continues forever. We show that at long times, the system along the longitudinal direction reaches a NESS, and we focus on this one-dimensional longitudinal  direction. 

We compute the density profile of the condensate in this NESS. This procedure also allows us to analytically explore other interesting observables, such as the distribution of the position of the rightmost particle of the condensate and the full counting statistics, i.e., the distribution of the number of particles in a segment $[-L, L]$. Our results illustrate the competition between the resetting-induced all-to-all attractions and the weakly repulsive interactions between the bosons in the condensate. 

The remainder of the paper is organized as follows. In \sref{sec:mp}, we introduce the harmonically trapped weakly interacting three-dimensional dilute Bose gas and discuss its one-dimensional Gross-Pitaevskii limit for a highly elongated cigar-shaped potential. We construct the Thomas-Fermi initial state, discuss its time evolution in the absence of a longitudinal trap, and define the stochastic resetting protocol. In \sref{s:no_resetting}, we first analyze the dynamics in the absence of resetting, showing that, following the quench of the longitudinal trapping frequency to zero, the condensate undergoes a self-similar free expansion governed by a time-dependent scaling factor. These results provide the basis for the analysis of the reset dynamics. In \sref{s:observables}, we characterize the resulting nonequilibrium steady state in the presence of resetting by computing the steady-state density profile, the statistics of the position of the rightmost particle of the condensate, and the full counting statistics of the number of particles contained within a finite interval. Finally, in \sref{sec:conc}, we summarize our main results and discuss future directions. Additional analytical and numerical details are relegated to the appendices.

\begin{figure}
    \centering
\includegraphics[width=0.95\textwidth]{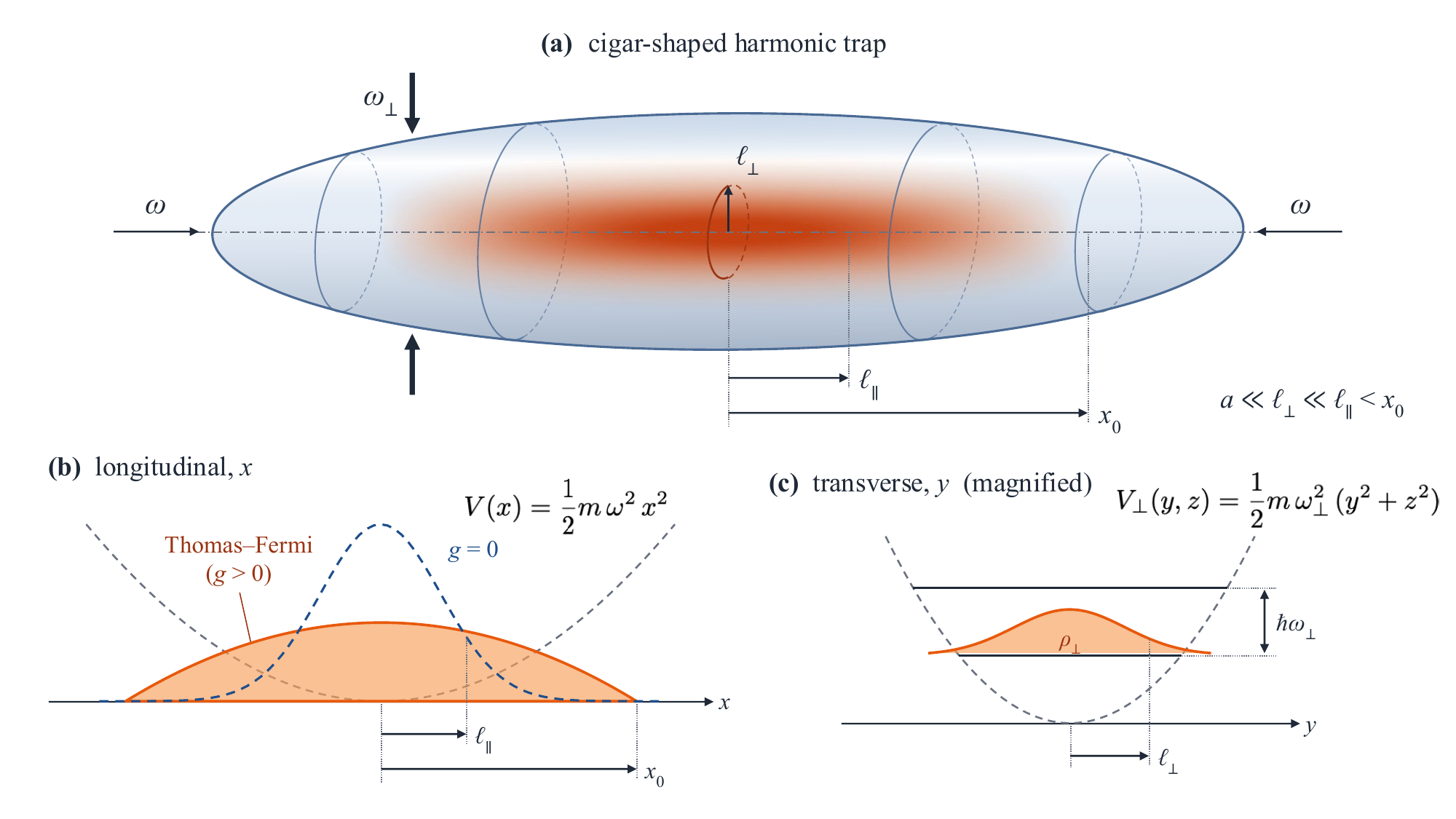}~
\caption{Schematic of the setup. (a) A gas of bosons in a cigar-shaped harmonic trap with strong transverse confinement $\omega_\perp$ (thick arrows) and weak longitudinal confinement $\omega$ (thin arrows), $\omega_\perp \gg \omega$. The shading shows the condensate density, $\propto \rho_{TF}(x)\,\rho_\perp(\boldsymbol{r}_\perp)$. Transversely, the gas is frozen in the harmonic-oscillator ground state of width $\ell_\perp$. Longitudinally, the repulsive interaction stretches it to the Thomas-Fermi length $x_0$ [Eq.~\eqref{eq:x0-2}], which exceeds the longitudinal oscillator length $\ell_\parallel$ since $x_0=\alpha\, C^{1/3}\ell_\parallel$ with $C\gg 1$ [Eq.~\eqref{eq:x0-1a}]. The oscillator lengths $\ell_\perp$ and $\ell_\parallel$ are defined in Eq.~\eqref{eq:lengths}, and the hierarchy of length scales follows from Eqs.~\eqref{eq:3d->1dcond} and \eqref{eq:x0-1a}. (b) Longitudinal density: the Thomas-Fermi profile $\rho_{TF}(x)$ [Eq.~\eqref{eq:TF_density}] (solid, $g>0$) compared with the noninteracting ($g=0$) ground state $e^{-x^2/\ell_\parallel^2}/(\sqrt{\pi}\,\ell_\parallel)$, i.e., the $C\to 0$ limit in Eq.~\eqref{eq:TDGPE2} (dashed), both normalized to unity. $\ell_\parallel$ is the half-width of the latter. The dashed grey curve is the longitudinal trapping potential $V(x)=\frac{1}{2}m\omega^2x^2$. (c) Transverse cut along $y$ (magnified) or equivalently $z$: the ground-state density along one of the transverse directions $\rho_\perp(y)=\exp(-y^2/\ell_\perp^2)/(\sqrt{\pi}\ell_\perp)$, whose half-width is $\ell_\perp$, drawn on the ground-state level $\hbar\omega_\perp$ of the transverse potential $V_\perp$. The gap $\hbar\omega_\perp$ to the first excited transverse level is much larger than the longitudinal energy per particle, which gives the last inequality in Eq.~\eqref{eq:3d->1dcond}. For illustration, $x_0=3\ell_\parallel$ in (a) and (b).}
    \label{fig:schem}
\end{figure}

\section{The model and the resetting protocol}
\label{sec:mp}

In this section, we define the model precisely and the protocol that we use for resetting. We consider a three-dimensional system of $N$ (large) bosons with weak, short-ranged repulsive pairwise interactions, confined in a highly elongated (cigar-shaped) harmonic trap (see Fig.~\ref{fig:schem}). 
We consider the limit where in the ground state of the three-dimensional system, the transverse motion is frozen to the ground state of the two-dimensional transverse harmonic oscillators, while the longitudinal density profile takes the Thomas-Fermi form~\cite{Dunjko_2, LS2001}. The reduction from the three dimensions to an effective one-dimensional 
Gross-Pitaevskii description  is well established~\cite{Dunjko_2, LS2001}. 
However, the relevant results are spread over works
that use different conventions and address broader questions. Our analysis relies on three successive approximations, namely the reduction to
one dimension, the mean-field (Gross-Pitaevskii) description, and the Thomas-Fermi
approximation. We therefore summarize this hierarchy of limits in
a self-contained form in~\sref{s:mod} and \sref{s:initial}. This also fixes the notation used in the rest of
the paper. Once the gas is prepared in this Thomas-Fermi state [\sref{s:initial}], the longitudinal trap is then switched off, and the Thomas-Fermi initial state is evolved [\sref{s:evolve}] till a random time $\tau$ picked from a PDF $p(\tau)=r\, e^{-r \tau}$. The process of starting from the Thomas-Fermi initial state and evolving in the absence of a trap till a random time is repeated endlessly. The detailed protocol is described in \sref{s:protocol}.

\subsection{Model}  
\label{s:mod}

We consider $N$ identical bosons of mass $m$  with a short-range repulsive interaction of s-wave scattering length $a$,  in an anisotropic harmonic trap in three dimensions, with frequency $\omega_\perp$ in the transverse directions ($\boldsymbol{r}_\perp$) and $\omega$ in the
longitudinal ($x$) direction. The characteristic length scales of the system are given by the transverse and longitudinal oscillator lengths, 
\begin{equation}
\ell_\perp = \sqrt{\frac{\hbar}{m\omega_\perp}}\, \qquad\text{and}\quad
\ell_\parallel = \sqrt{\frac{\hbar}{m\omega}}\, ,
\label{eq:lengths}
\end{equation}
respectively. The effective one dimensional density of the bosons along the $x$-direction is defined as $\rho(x) = \int \rho_\text{3D} (\boldsymbol{r})\, d \boldsymbol{r}_\perp$, where 
$ \rho_\text{3D} (\boldsymbol{r})$ denotes the density of the gas in three dimensions at the position vector $\boldsymbol{r}$. 
{With $\rho(x)$ normalized to unity, $n(x)=N\rho(x)$ is the local one-dimensional number
density. In a box of length $L$ without a longitudinal potential, $n(x)=N/L$ is uniform
and directly characterizes the density of the gas. In a confining potential, however,
$n(x)$ varies with position, and a single characteristic density requires an average.
A plain spatial average would depend on an arbitrary choice of the averaging length,
since a trapped cloud has no sharp size in general. Instead, we average $n(x)$ with the
weight $\rho(x)$, i.e., over the positions of the particles, which gives the density
experienced by a typical particle~\cite{LS2001},
\begin{equation}
   \bar{n} = \int_{-\infty}^\infty n(x)\, \rho(x)\, dx = N \int_{-\infty}^\infty \rho^2(x)\, dx \, .
\label{eq:nbar}
\end{equation}
For a uniform gas in a box, Eq.~\eqref{eq:nbar} reduces to $\bar n=N/L$. For a
non-interacting gas in the harmonic trap, $\bar n=N/(\sqrt{2\pi}\,\ell_\parallel)$.
Repulsive interactions broaden the cloud to a larger size $x_0$, which depends on $N$ and
on the interaction strength, so that $\bar n\propto N/x_0$. Since $\rho(x)$, and hence
$x_0$, is determined only later (see \sref{s:initial}), $\bar n$ is not known at this
stage.}

It has been shown that~\cite{Dunjko_2, LS2001}, in the limit
\begin{equation}
 N\to\infty \quad\text{and}~ \quad a\ll \ell_\perp \ll \ell_\parallel \quad\text{such that}\quad   \ell_\perp^2\,\bar{n}\, \min\{\bar{n}, a/\ell_\perp^2\} \ll  1  ,
  \label{eq:3d->1dcond}
\end{equation}
in the ground state of the three-dimensional system, the transverse direction is frozen to the ground state of the $N$ non-interacting two-dimensional harmonic oscillators with energy per particle $\hbar \omega_\perp$ and a Gaussian density profile $\rho_\perp(\boldsymbol{r}_\perp) = \exp(-r_\perp^2/\ell_\perp^2)/(\pi \ell_\perp^2)$. The ground state energy and the density profile in the longitudinal direction (which is of our interest here) can be effectively described by a  one-dimensional Hamiltonian~\cite{Dunjko_2, LS2001}
\begin{equation}
\label{eq:H(w)_field}
    \hat{\mathcal{H}}(\omega) = \int_{-\infty}^{\infty} dx \,  \hat{\Psi}^{\dagger}(x) \left[ -\frac{\hbar^2}{2m} \frac{\partial^2}{\partial x^2}  + \frac{1}{2} m \omega^2 x^2  \right] \hat{\Psi}(x) + \frac{g}{2}\int_{-\infty}^{\infty} dx \, \hat{\Psi}^{\dagger}(x)\hat{\Psi}^{\dagger}(x)\hat{\Psi}(x) \hat{\Psi}(x) \, ,
\end{equation}
where the repulsive  pair coupling strength $g$ is given by~\cite{Dunjko_2}
\begin{equation}
    g= \frac{2 \hbar^2 a}{m \ell_\perp^2}=2 \hbar\, a\, \omega_\perp \, .
    \label{eq:g}
\end{equation}
The last inequality in Eq.~\eqref{eq:3d->1dcond} follows from the condition that the average energy per particle in the longitudinal direction is much smaller than the gap $\hbar\omega_\perp$ between the ground state and the first excited state in the transverse direction~\cite{LS2001}. Note that the condition in Eq.~\eqref{eq:3d->1dcond} implies that the gas is dilute in three dimensions, i.e., $ a^3 n_\text{3D}   \sim a^3 \bar{n} /\ell_\perp^2 \ll 1$, but need not be dilute in one dimension~\cite{LS2001}, as we also see below.

The field operators in Eq.~\eqref{eq:H(w)_field} satisfy the canonical bosonic
commutation relations 
\begin{equation}
  [\hat{\Psi}(x), \hat{\Psi}^{\dagger}(x')] = \delta(x-x'), \quad\text{and} \quad [\hat{\Psi}(x), \hat{\Psi}(x')] = [\hat{\Psi}^{\dagger}(x), \hat{\Psi}^{\dagger}(x')]=0.  
  \label{eq:bose-com}
\end{equation} 
By using the Heisenberg equations of motion $i \hbar \, \partial_t\hat{\Psi}(x,t) = [\hat{\Psi}(x,t), \hat{\mathcal{H}}(\omega)]$, along with the bosonic commutation relations in Eq.~\eqref{eq:bose-com}, one finds the equation of motion for the field operator as
\begin{equation}
\label{eq:sq_ham}
    i \hbar \, \frac{\partial}{\partial t} \hat{\Psi}(x,t) = -\frac{\hbar^2}{2m} \frac{\partial^2}{\partial x^2}  \,  \hat{\Psi}(x,t) + \frac{1}{2} m \omega^2 x^2 \hat{\Psi}(x,t) + g [\hat{\Psi}^\dagger(x,t) \hat{\Psi}(x,t)] \hat{\Psi}(x,t) \, .
\end{equation}
It is useful to express the repulsive pair coupling strength $g$ given in Eq.~\eqref{eq:g}, in terms of  the dimensionless Lieb-Liniger parameter
\begin{equation}
   \gamma = \frac{m g}{\hbar^2 \bar{n}} = \frac{2 a }{\ell_\perp^2\, \bar{n}} \, ,
  \label{eq:gamma}
\end{equation}
where $\bar{n}$ is defined in Eq.~\eqref{eq:nbar} and we have used the expression of $g$ from Eq.~\eqref{eq:g} in the last step.  
In the weak coupling limit $\gamma \ll 1$, which also  corresponds to the high one-dimensional density limit,
 one can use the Bogoliubov approximation~\cite{PS2016, Dalfovo_1} and expand the field operator as
$\hat{\Psi}(x,t) = \sqrt{N} \psi(x,t)  + \delta\hat{\psi}(x,t) $, where $\psi(x,t)$ is a classical (complex) field. 
By neglecting the small perturbation $\delta \hat{\psi}(x,t)$, and therefore, substituting $\hat{\Psi}(x,t) = \sqrt{N} \psi(x,t)$ in Eq.~\eqref{eq:sq_ham}, one gets the GPE~\cite{PS2016, Pitaevskii1961, Gross1963, LS2001, ELS2007, Dalfovo_1} for
the condensate wavefunction $\psi(x,t)$ as 
\begin{equation}
	i\hbar\frac{\partial}{\partial t}\psi(x,t)
	=\Biggl[-\frac{\hbar^2}{2m}\frac{\partial^2}{\partial x^2}+  \frac{1}{2}m\,\omega^2 x^2 + g N |\psi(x,t)|^{2} \Biggr]\psi(x,t) \, .
 \label{eq:TDGPE} 
\end{equation}
In this effective description, the repulsive pair interaction between the particles is captured by the nonlinear term in Eq.~\eqref{eq:TDGPE}. 
The average density of bosons in the condensate is given by 
\begin{equation}
    \rho(x,t)=|\psi(x,t)|^2\, ,
    \label{eq:density100}
\end{equation}
where $\psi(x,t)$ is obtained from the solution of Eq.~\eqref{eq:TDGPE}. It is important to note that, here, the density refers to the fractional density [and not the number density $n(x)$], which is normalized to unity (and not $N$), i.e., $\int_{-\infty}^\infty\rho(x,t)\, dx =1$. The term $gN|\psi(x,t)|^2$ in Eq.~\eqref{eq:TDGPE} is the mean-field potential that one particle feels from all the others. Since each pair of particles is counted twice (once from each particle's side), taking the expectation value of  $gN|\psi(x,t)|^2$ with respect to $\psi(x,t)$ gives twice the interaction energy per particle,  
\begin{equation}
  2\frac{E_\text{int}}{N} = \int_{-\infty}^\infty \psi^*(x,t) gN |\psi(x,t)|^2\, \psi(x,t)\, dx = g N  \int_{-\infty}^\infty \rho^2(x,t)\, dx = g\bar{n}(t)\, . 
  \label{eq:Eint}
\end{equation}
Therefore,  the interaction energy per particle is given by $E_\text{int}/N = g\bar{n}(t)/2$.

To study the ground state properties of the GPE in Eq.~\eqref{eq:TDGPE}, it is useful to express the equation in dimensionless form.  We are interested in the case where 
the harmonic confining potential term and the nonlinear term arising from the repulsive pairwise  interaction in Eq.~\eqref{eq:TDGPE}  are of the same order. Since $|\psi(x)|^2$ is normalized to unity, substituting the scaling form 
\begin{equation}
    \psi(x,t)= \frac{1}{\sqrt{x_0}}\, \chi(x/x_0, t)\, ,
    \label{eq:chi}
\end{equation}
in Eq.~\eqref{eq:TDGPE}, and comparing the second and third terms on the right-hand side, we get
\begin{equation}
    x_0  =\alpha\, C^{1/3}\, \ell_\parallel, \quad\text{where}~~ C = \frac{m g N\ell_\parallel}{\hbar^2} =\frac{2 N a \ell_\parallel}{\ell_\perp^2} \, ,
    \label{eq:x0-1a}
\end{equation}
and $\alpha$ is an arbitrary numerical constant of $O(1)$, which we fix later.  We note that $x_0$ has the dimension of length whereas $C$ is dimensionless.
Using Eqs.~\eqref{eq:chi} and \eqref{eq:x0-1a}, we can rewrite  Eq.~\eqref{eq:TDGPE} in dimensionless form
\begin{equation}
    \frac{i}{\omega} \frac{\partial}{\partial t} \chi(y,t) = - \frac{1}{2 \alpha^2\,C^{2/3}}\, \frac{\partial^2 }{\partial y^2} \chi(y,t) + C^{2/3} \left[\frac{1}{2}  \alpha^2 y^2 + \frac{1}{\alpha}|\chi(y,t)|^2 \right]\chi(y,t)\, ,
    \label{eq:TDGPE1a}
\end{equation}
where $C$ is $O(1)$ or larger.

We remark that the scaling form in Eq.~\eqref{eq:chi} is not appropriate in the limit $C\to 0$,  since from Eq.~\eqref{eq:x0-1a}, $x_0\to 0$ in that  limit. Consequently, Eq.~\eqref{eq:TDGPE1a} is not suitable to  take the limit $C\to 0$. In this limit, one needs to scale $x$ with $\ell_\parallel$, i.e., $\psi(x,t)=\chi_0(x/\ell_\parallel, t)/\sqrt{\ell_\parallel}$, in which case  Eq.~\eqref{eq:TDGPE} can be written in the dimensionless form as
\begin{equation}
    \frac{i}{\omega} \frac{\partial}{\partial t} \chi_0(y,t) = \left[- \frac{1}{2}\, \frac{\partial^2 }{\partial y^2}  + \frac{1}{2}   y^2 + C|\chi_0(y,t)|^2 \right]\chi_0(y,t)\, .
    \label{eq:TDGPE2}   
\end{equation}
Therefore, in the $C\to 0$ limit, the mean-field description in Eq.~\eqref{eq:TDGPE2} becomes that of an ideal gas~\cite{LS2001}, which we will not discuss in this work. Equation~\eqref{eq:TDGPE1a} is the suitable one for our purpose.

\subsection{Preparation of the initial state} 
\label{s:initial}

In the trapping potential, the GPE in Eq.~\eqref{eq:TDGPE} admits a solution of the type 
\begin{equation}
\label{eq:phix}
    \psi(x,t) = e^{-i \mu t/\hbar} \phi(x) \,, 
\end{equation} 
where $\phi(x)$ satisfies the time-independent equation
\begin{equation}
    \mu \, \phi(x) = \Biggl[-\frac{\hbar^2}{2m}\frac{d^2}{d x^2}+\frac{1}{2}m\,\omega^2 x^2+  g N |\phi(x)|^{2} \Biggr]\phi(x) \,  ,
 \label{eq:GPE}
\end{equation}
where $\mu$ is yet to be determined. For this class of solutions, the density $\rho(x,t)=|\psi(x,t)|^2=|\phi(x)|^2$ remains invariant in time. Multiplying Eq.~\eqref{eq:GPE}  by $\phi^*(x)$ and integrating over $x$ gives
\begin{equation}
    \mu = \frac{E_\text{kin}+ E_\text{trap}}{N} + g\bar{n} = \frac{E+E_\text{int}}{N},
\end{equation}
where $E_\text{kin}/N$ and $E_\text{trap}/N$ are the average kinetic and trap energy per particle respectively.  The average interaction energy per particle $E_\text{int}/N=g\bar{n}/2$ is obtained in Eq.~\eqref{eq:Eint}. Finally  $E/N=(E_\text{kin}+E_\text{trap} + E_\text{int})/N$ is the average total energy per particle.

Equation~\eqref{eq:GPE} is still a nonlinear differential equation and it is very hard to solve it exactly. It follows from Eqs.~\eqref{eq:chi} and \eqref{eq:phix} that 
\begin{equation}
    \phi(x) = \frac{1}{\sqrt{x_0}}\, \Phi(x/x_0)\, \quad \text{and}\quad \chi(y,t) = e^{-i\mu t/\hbar} \,\Phi(y),  
\end{equation}
where, from Eq.~\eqref{eq:TDGPE1a}, $\Phi(y)$ satisfies the time-independent equation
\begin{equation}
    \frac{\mu}{\hbar\omega} \, \Phi(y) = 
    - \frac{1}{2 \alpha^2\,C^{2/3}}\,  \Phi''(y) + C^{2/3} \left[\frac{1}{2}  \alpha^2 y^2  + \frac{1}{\alpha}|\Phi(y)|^2 \right]\Phi(y)\, 
    \label{eq:GPE1}
\end{equation}
It can be easily verified that Eqs.~\eqref{eq:GPE} and \eqref{eq:GPE1} are equivalent. As mentioned above, the nonlinear equation~\eref{eq:GPE1} is hard to solve for arbitrary $C$. Hence, we take the limit $C \gg 1$, which is known as the Thomas-Fermi (TF) approximation~\cite{BP1996, CD1996}.  In this limit, the kinetic term $\Phi''(y)$ drops out of Eq.~\eqref{eq:GPE1}, and we find that $ \Phi(y)$ is given by the semi-circular form
\begin{equation}
    \Phi(y) = \sqrt{\frac{\alpha \mu}{\hbar \omega \,C^{2/3}} - \frac{\alpha^3 y^2}{2} }\, \Theta(y_0-|y|) , \quad \text{with}\quad y_0 = \frac{1}{\alpha \, C^{1/3}}\, \sqrt{\frac{2 \mu}{\hbar\omega}}\, ,
    \label{eq:y0}
\end{equation}
where $\Theta(z)$ is the Heaviside step function, i.e., $\Theta(z)=1$ for $z>0$, and zero otherwise. The normalization condition $\int_{-\infty}^\infty |\Phi(y)|^2\, dy=1$ determines $\mu$ as a function of $C$ as
\begin{equation}
    \mu(C) = \frac{1}{2}\left(\frac{3}{2}\right)^{2/3}  \, \hbar\omega \, C^{2/3}.
    \label{eq:mu}
\end{equation}
We now substitute $\mu(C)$ from Eq.~\eqref{eq:mu} in $y_0$ in Eq.~\eqref{eq:y0},  subsequently set $y_0=1$, and determine $\alpha$ as
\begin{equation}
\label{eq:alpha}
    \alpha = \left(\frac{3}{2}\right)^{1/3} = 1.144714\dots
\end{equation}
Substituting $\mu$ and $\alpha$ from Eqs.~\eqref{eq:mu} and \eqref{eq:alpha} in Eq.~\eqref{eq:y0} gives
\begin{equation}
    \Phi (y)  = \sqrt{\frac{3}{4} (1-y^2)} \,\,  \Theta(1- |y|) \, .
    \label{eq:phiy}
\end{equation}

In summary, in the Thomas-Fermi limit $C\gg 1$ of the GPE, which corresponds to $N^{-2} \ll \gamma \ll 1$ , the function $\phi(x)$ in Eq.~\eqref{eq:phix} has the scaling form
\begin{equation}
\label{eq:TF_wavefn}
    \phi(x) \approx \phi_{TF}(x) = \frac{1}{\sqrt{x_0}} \, 
    \Phi {\left( \frac{x}{x_0} \right)},
\end{equation}
where the scaling function $\Phi(y)$ is given by Eq.~\eqref{eq:phiy}, and $x_0$ is given in Eq.~\eqref{eq:x0-1a} with $\alpha$ in Eq.~\eqref{eq:alpha}, i.e., 
\begin{equation}
    x_0 = \left(\frac{3   g N}{2 m \omega^2}\right)^{1/3} .
    \label{eq:x0-2}
\end{equation}

We now define the corresponding TF density $\rho_{TF}(x) = | \phi_{TF}(x)|^2$, which has the scaling form
\begin{equation} 
\label{eq:TF_density}
    \rho_{TF}(x) = \frac{1}{x_0} \, R{\left( \frac{x}{x_0} \right)} \quad \text{with} \quad R(z) = \left|\Phi{\left(z \right)} \right|^2 = \frac{3}{4} \, (1-z^2) \, \Theta(1-z^2) . 
\end{equation}
The state
\begin{equation}
    \psi(x,0)=\phi_{TF} (x),
    \label{eq:initial_state}
\end{equation}
where $\phi_{TF}(x)$ is given in Eq.~\eqref{eq:TF_wavefn}, is going to be our initial state. 

Using the Thomas-Fermi density $\rho_{TF}(x)$ in Eq.~\eqref{eq:TF_density}, one can easily compute the one-dimensional mean number density as $\bar{n} = 3N/(5 x_0)$, and subsequently, $\gamma = (5/3) (3/2)^{1/3} C^{4/3}/N^2\simeq 1.91 C^{4/3}/N^2$, and the condition $\gamma\ll 1$ yields $C \ll N^{3/2}$. In other words, the condition $N^{-2} \ll \gamma \ll 1$ for the Thomas-Fermi approximation  can also be cast  as $ 1\ll C \ll N^{3/2}$.

\begin{figure}
    \centering
    \includegraphics[width=.49\textwidth]{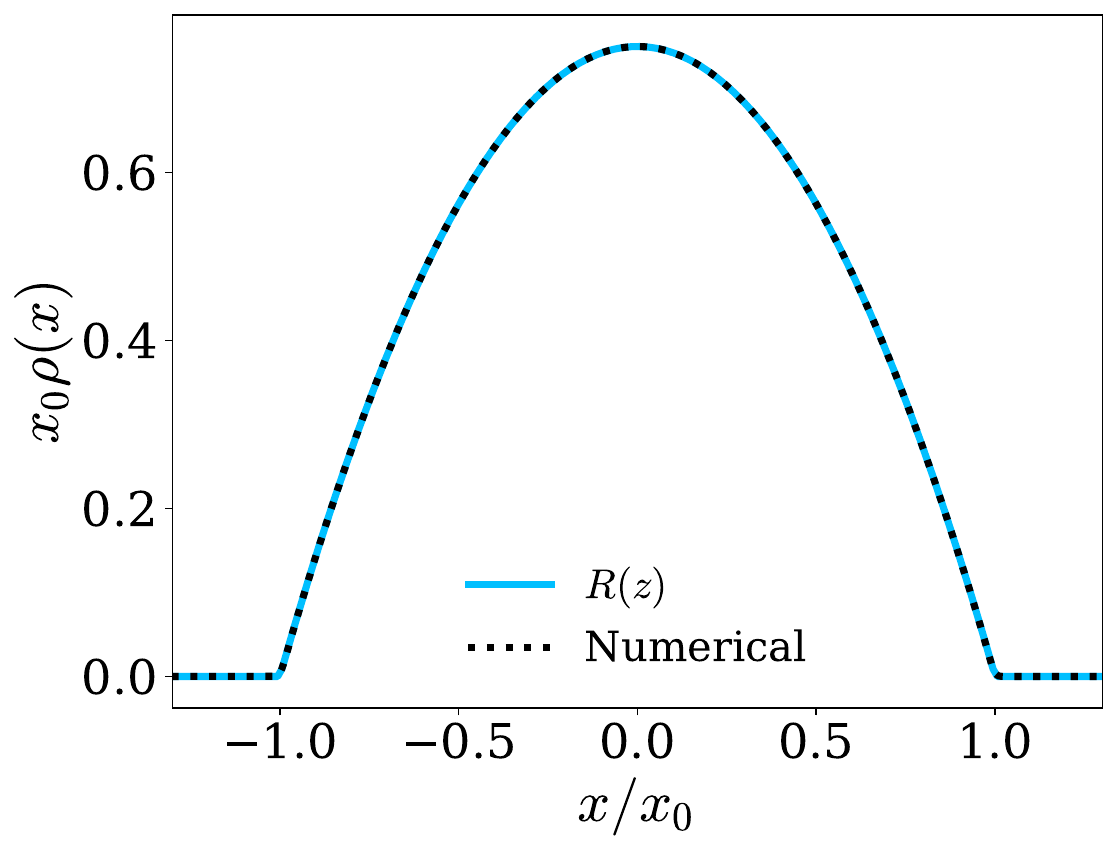}~
\includegraphics[width=.49\textwidth]{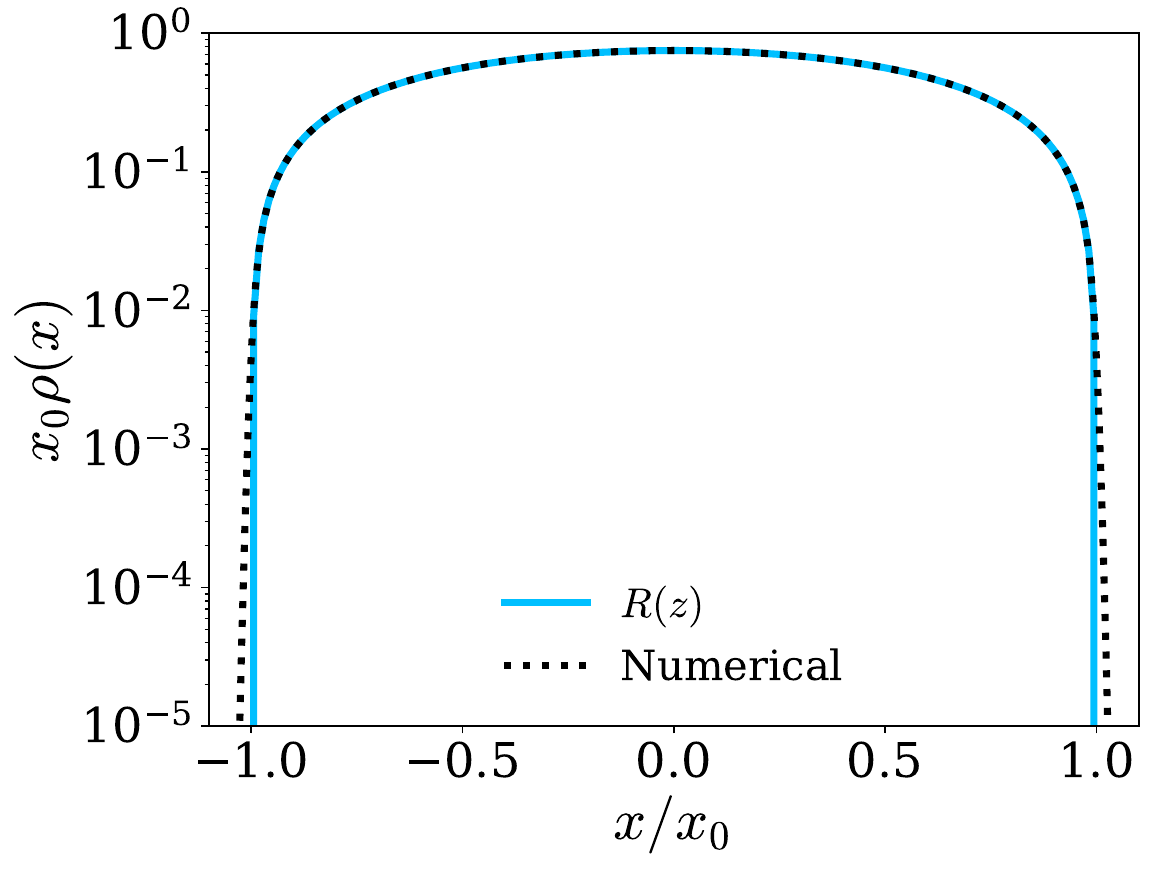}
\caption{Condensate density of weakly interacting bosons in harmonic confinement, given by the Thomas-Fermi (TF) solution of the time-independent GPE in Eq.~\eqref{eq:GPE}.
The dashed black lines are obtained by numerically solving Eq.~\eqref{eq:GPE} and the solid blue line is the scaled TF density $R(z) \equiv \,  x_0 \rho(x)$ with $z=x/x_0$ in Eq.~\eqref{eq:TF_density},  in (left) linear-linear scale  and (right) log-linear scale. The numerical evaluation of the GPE involves the  evolution of the time-dependent GPE [Eq.~\eqref{eq:TDGPE1}] in imaginary time as detailed in \ref{A:sim_gs}. The parameters used are $N = 5 \times 10^5$, $m=0.4$, $\omega=0.1$, $g=0.01$. } 
    \label{fig:gs}
\end{figure}

In Fig.~\ref{fig:gs},  we  plot the Thomas-Fermi density profile along with the numerically obtained solution of the GPE in Eq.~\eqref{eq:GPE} (details in~\ref{A:sim_gs}), and we find an excellent agreement between the two. The log-linear plot in Fig.~\ref{fig:gs} (right) shows that the numerical solution of the GPE closely follows the TF approximation everywhere except near the edge $x \approx x_0$, where there is a negligible deviation.

In summary, starting from the three-dimensional dilute Bose gas with repulsive short-ranged pair interaction in an anisotropic harmonic trap, we go through the following three main approximation steps: 
\begin{enumerate}
    \item The condition in Eq.~\eqref{eq:3d->1dcond} enables the ground state of the gas to be effectively treated as a one-dimensional Bose gas with delta-function repulsive interaction along the longitudinal direction, whereas the gas in the transverse direction settles down to the ground state of noninteracting two-dimensional harmonic oscillators.  

    \item In the weak coupling limit $\gamma \ll 1$ [defined in Eq.~\eqref{eq:gamma}], the density of the one-dimensional gas in the ground state can be described by the GPE. In this  limit, while the three-dimensional gas is dilute, the effective one-dimensional gas in the longitudinal direction is highly dense. Moreover, for $\gamma\ll 1$, from Eq.~\eqref{eq:gamma}, $ a/\ell_\perp^2 \ll \bar{n}$. Therefore, using $\min\{\bar{n}, a/\ell_\perp^2\}=a/\ell_\perp^2$,   the last inequality in Eq.~\eqref{eq:3d->1dcond} simplifies to $ a\bar{n} \ll 1\, $. 

    \item In the limit $N^{-2} \ll \gamma \ll 1$, the ground state density profile along the longitudinal direction can be approximated by the Thomas-Fermi form [Eq.~\eqref{eq:TF_density}]. 
\end{enumerate}

\subsection{Time evolution} 
\label{s:evolve}

The time evolution of the Bose-Einstein condensate in a time dependent harmonic potential starting from the Thomas-Fermi initial state in Eq.~\eqref{eq:initial_state} has been investigated earlier~\cite{CD1996,Kagan96,exp_free}. Closely related self-similar scaling dynamics of a Thomas--Fermi quasi-one-dimensional Bose gas have been observed experimentally under
time-dependent longitudinal confinement~\cite{Rohringer2015}. Here, we consider the particular case in which the longitudinal trap
is switched off completely (the transverse confinement
stays on) and monitor its evolution as a function of time.
We show that there is a self-similar time-dependent solution that is
exact within the Thomas–Fermi (hydrodynamic) approximation at all times $t$. 
Starting from the initial state in Eq.~\eqref{eq:initial_state},  the GPE in Eq.~\eqref{eq:TDGPE} in the absence of the longitudinal trap ($\omega=0$)
evolves as
\begin{equation}
	i\hbar\frac{\partial}{\partial t}\psi(x,t)
	=\Biggl[-\frac{\hbar^2}{2m}\frac{\partial^2}{\partial x^2} + g N |\psi(x,t)|^{2} \Biggr]\psi(x,t) \, .
 \label{eq:TDGPE1} 
\end{equation}
To solve this equation, we first write 
\begin{equation}
\label{eq:cov}
    \psi(x,t) =\sqrt{\rho(x,t)}\, e^{\frac{i}{\hbar} S(x,t)}\,\quad\text{with}\quad  \rho(x,0) = |\phi_{TF}(x)|^2 \quad \text{and} \quad S(x,0)=0.
\end{equation}
Substituting this form in Eq.~\eqref{eq:TDGPE1}, and equating the imaginary parts, we get the continuity equation
\begin{equation}
\label{eq:cont}
    \frac{\partial \rho}{\partial t} +\frac{\partial}{\partial x}(\rho v) = 0\,,
\end{equation}
where the velocity $v(x,t)$ is given by 
\begin{equation}
\label{eq:vel}
    v(x,t)=\frac{1}{m}\frac{\partial S}{\partial x}\,\quad \text{with} \quad v(x,0) = 0\,.
\end{equation}
Similarly, equating the real parts,  and subsequently,  taking a derivative with respect to $x$, we get the Euler equation
\begin{equation}
\label{eq:euler}
    m\frac{\partial v}{\partial t} +\frac{\partial}{\partial x}\Big(\frac{1}{2}mv^2+ g\, N \,\rho+ V_Q \Big) = 0\, \quad \text{with} \quad V_Q = -\frac{\hbar^2}{2m} \frac{1}{\sqrt{\rho}} \frac{\partial^2 \sqrt{\rho}}{\partial x^2}\, .
\end{equation}
The term $V_Q$ is the so-called quantum pressure or Madelung term.

To solve the coupled continuity and  Euler equations, \eqref{eq:cont} and \eqref{eq:euler} respectively,  inspired by the scaling form of the initial condition $\rho(x,0) = \rho_{TF}(x)$ given in Eq.~\eqref{eq:initial_state}, we make the following ansatz 
\begin{equation}
\label{eq:rscale}
    \rho(x,t) =\frac{1}{x_0\lambda(t)} R\left(\frac{x}{x_0\lambda(t)}\right) \, \quad \text{with}\quad \lambda(0)=1\,,
\end{equation}
where $R(z)$ is given by Eq.~\eqref{eq:TF_density}. In other words, we seek for a self-similar scaling solution, where the expansion of the gas is dictated solely by the function $\lambda(t)$. The ansatz in Eq.~\eqref{eq:rscale} should satisfy both Eqs.~\eqref{eq:cont} and \eqref{eq:euler}.
First, substituting this ansatz in the continuity equation~\eqref{eq:cont} and using the initial condition $v(x,0)=0$, we get 
\begin{equation}
\label{eq:v1}
    v(x,t) = \frac{\dot{\lambda}(t)}{\lambda(t)}x\,  \quad \text{with}\quad \dot{\lambda}(0)=0\,.
\end{equation}
We next use the ansatz for the density in Eq.~\eqref{eq:rscale} and the solution $v(x,t)$ from Eq.~\eqref{eq:v1} in Eq.~\eqref{eq:euler}. Since from the expression of the edge of the support $x_0$ from Eq.~\eqref{eq:x0-2}, the space scales as $x\sim N^{1/3}$, it is evident from Eq.~\eqref{eq:v1} that $v(x,t) \sim x/t \sim N^{1/3}$ and $\rho(x,t)\sim 1/x \sim N^{-1/3}$. Therefore, it is easy to see that all the terms except the $V_Q$ term in Eq.~\eqref{eq:euler} scale as $O(N^{1/3})$ for large $N$. The $V_Q$, term on the other hand, scales as  $O(1/N)$, and hence can be neglected for large $N$. Substituting Eq.~\eqref{eq:v1} in the Euler equation~\eqref{eq:euler}, ignoring $V_Q$, and using Eq.~\eqref{eq:rscale} along with the expression of $R(z)$ from Eq.~\eqref{eq:TF_density}, we
get
\begin{equation}
     \frac{d^2 \lambda(t)}{dt^2} = \frac{\omega^2}{\lambda^2(t)} \quad\quad\text{with }\quad\quad
     \lambda(0) =1 \quad\text{and}\quad
     \dot{\lambda}(0) = 0\,.
     \label{eq:lambda(t)}
\end{equation}
It is interesting to note from this equation that $\lambda(t)$ only depends on the initial trap frequency $\omega$ and is completely independent of the interaction strength $g$.
The solution of $\lambda(t)$ is discussed in \sref{s:no_resetting}. 

\subsection{The quench protocol} 
\label{s:protocol}

The quench protocol that we use in our work is as follows: 

\begin{enumerate}
    \item Initial step: Prepare the gas in the Thomas-Fermi state [Eq.~\eqref{eq:initial_state}] of the harmonically trapped GPE with trap frequency $\omega$ [Eq.~\eqref{eq:GPE}].

    \item Free expansion: At $t=0$, suddenly switch off the harmonic trap by quenching the frequency $\omega\to 0$. The gas then evolves according to the free GPE~\eqref{eq:TDGPE1} for a random time $\tau$, independently drawn from the exponential distribution $p(\tau)=r e^{-r\tau}$.

    \item Resetting: At time $\tau$, switch the harmonic trap back on and instantaneously cool the gas to the initial Thomas-Fermi state [Eq.~\eqref{eq:initial_state}]. This step resets the system to its initial configuration.

    \item Repeat the alternating free-expansion and resetting steps indefinitely, drawing a new independent waiting time $\tau$ from $p(\tau)$ after each reset.
\end{enumerate}

\section{Free expansion in the absence of resetting}
\label{s:no_resetting}

From the previous section, it is clear that, in order to obtain an explicit form of the density profile $\rho(x,t)$ in Eq.~\eref{eq:rscale} during the free-expansion phase of our protocol, starting from the Thomas-Fermi initial state [Eq.~\eqref{eq:initial_state}], we need to solve for $\lambda(t)$ in Eq.~\eref{eq:lambda(t)}.  In this section, we discuss free expansion in the absence of resetting. This discussion will also establish the results needed for analyzing the system under stochastic resetting in the following sections.

\begin{figure}
    \centering
    \includegraphics[width=.49\textwidth]{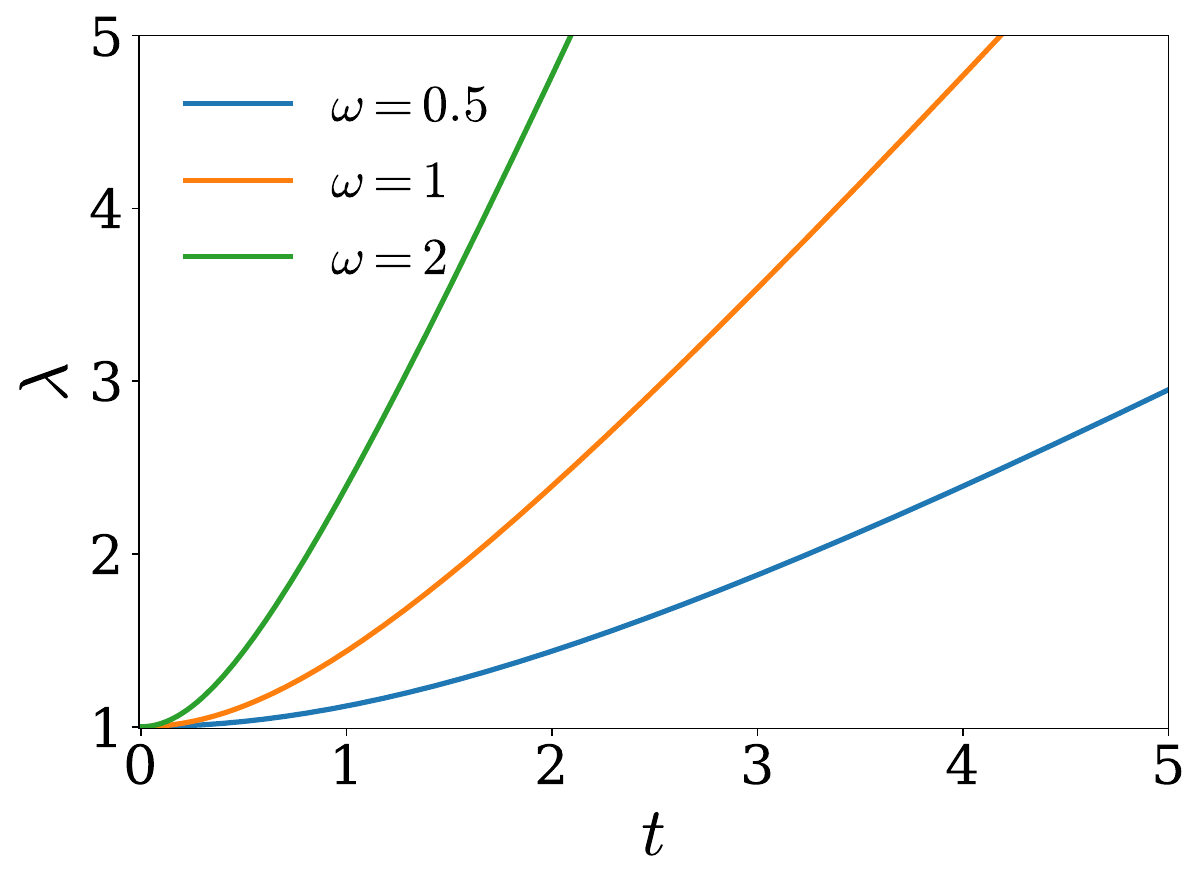}~
\includegraphics[width=.49\textwidth]{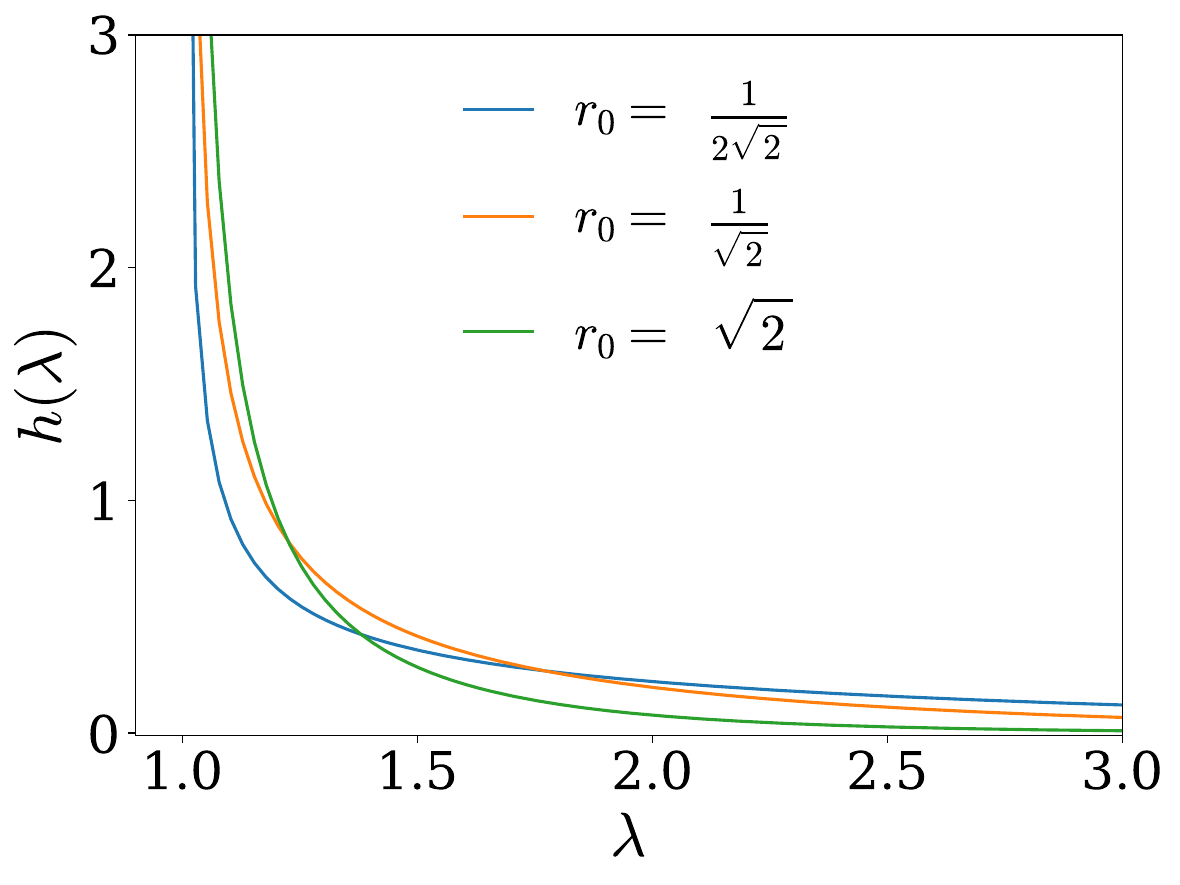}
    \caption{ (Left) The plot of $\lambda$ as a function of $t$ for various values of $ \omega$. The plot is obtained by finding the roots of Eq.~\eqref{eq:lambda} numerically. (Right)
    Plot of $h(\lambda)$ given in Eq.~\eqref{eq:h_lam} for various values of $r_0 = r/(\sqrt{2}\omega)$.}  
    \label{fig:lam}
\end{figure}

Multiplying Eq.~\eqref{eq:lambda(t)} by $\dot{\lambda}$ on both sides gives
\begin{equation}
    \frac{1}{2}\frac{d}{dt} \left(\dot{\lambda}^2\right) = - \omega^2 \frac{d}{dt} \left(\frac{1}{\lambda}\right).
\end{equation}
Integrating this equation and using the initial conditions in Eq.~\eqref{eq:lambda(t)} yields
\begin{equation}
\frac{d\lambda}{dt} = \sqrt{2} \, \omega \, \sqrt{1-\frac{1}{\lambda}}.
   \label{eq:beta1a}
\end{equation}
We remark that in Eq.~\eqref{eq:beta1a}, we chose the positive root with $\dot{\lambda}(t)>0$. This is because 
\begin{equation}
    \dot{\lambda}(t) = \int_0^t \ddot{\lambda}(t')\, dt' = \int_0^t \frac{\omega^2}{\lambda^2(t')}\, dt' >0.
\end{equation}
This is natural since $\lambda(t)$ in Eq.~\eqref{eq:rscale}, represents the boundary of the gas. Via free evolution we expect the gas to expand and not to contract due to the repulsive interaction. This then dictates the choice  $\dot{\lambda}(t)>0$. Given the initial condition $\lambda(0)=1$ and the fact that $\dot{\lambda}(t)>0$ for all $t$, it follows that $\lambda(t)\ge 1$ for all $t\geq0$.

Integrating Eq.~\eqref{eq:beta1a} with the initial condition $\lambda(0) = 1$, gives $\lambda(t)$ implicitly via the inverse function
\begin{equation}
\label{eq:lambda}
 t(\lambda)  =
 \frac{1}{\sqrt{2}\, \omega }  \left[   \sqrt{\lambda(\lambda-1)} \, + \ln \left(\sqrt{\lambda}+\sqrt{\lambda -1}\right) \right] \, .
\end{equation}
Taking a derivative of Eq.~\eqref{eq:lambda} with respect to $\lambda$, we get 
\begin{equation}
    t'(\lambda) = \frac{1}{\sqrt{2}\, \omega}\, \sqrt{\frac{\lambda}{\lambda-1}} \, > 0 \quad\text{for}\quad \lambda>1.
\end{equation}
Therefore, $t(\lambda)$ is a monotonic function of $\lambda$. When $\lambda\to 1$, we get $t\to 0$, whereas $t\to \infty$ as $\lambda\to \infty$. Thus, Eq.~\eqref{eq:lambda} has a unique inverse $\lambda(t)$, which is shown in \fref{fig:lam}.

\begin{figure}
    \centering
    \includegraphics[width=.49\textwidth]{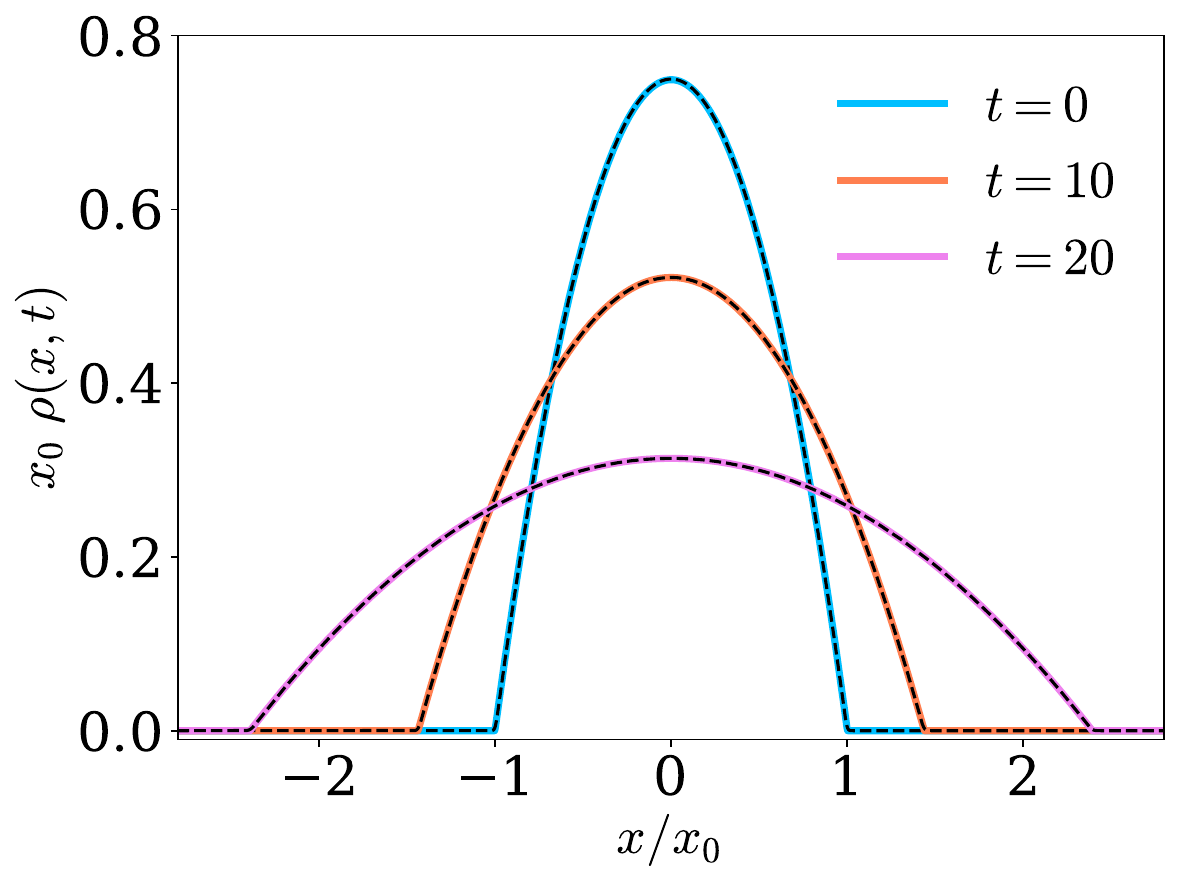}~\includegraphics[width=.49\textwidth]{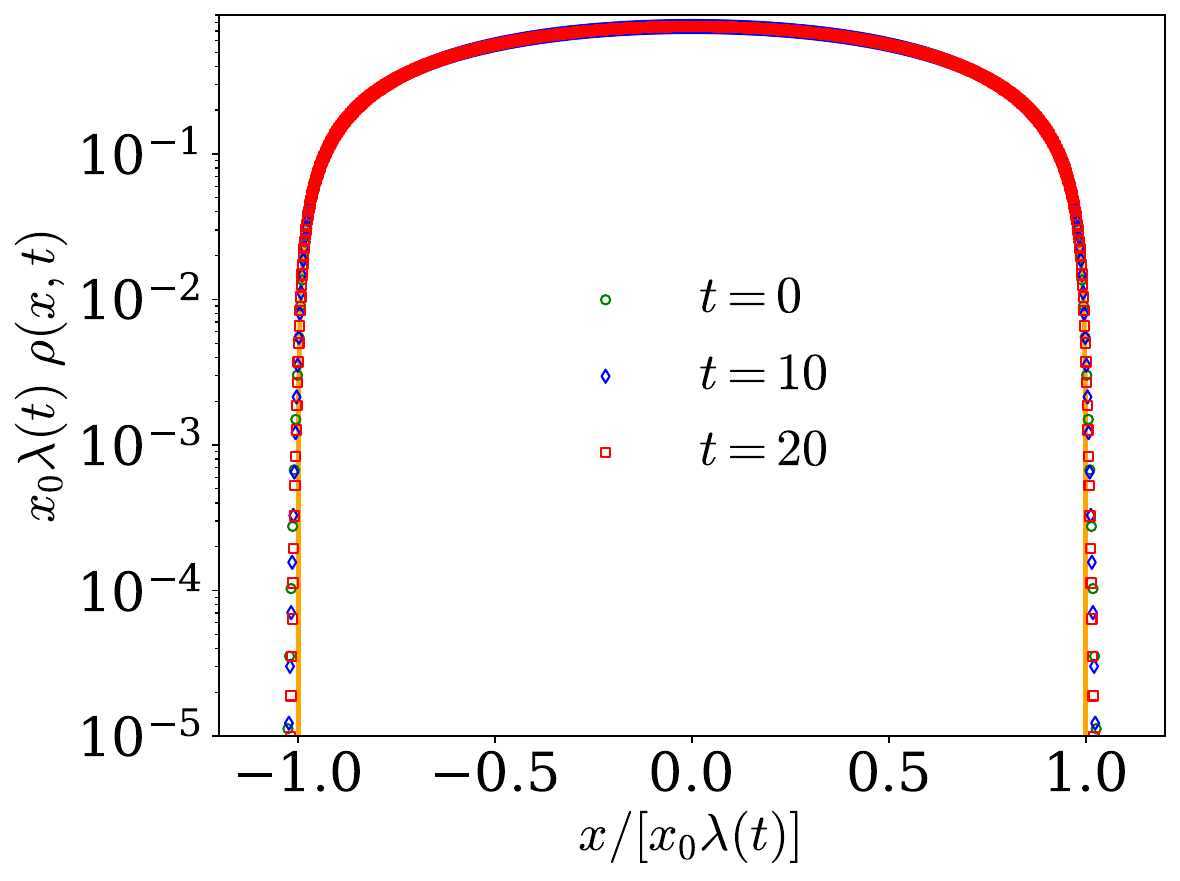}
\caption{Dynamics of the condensate density of weakly interacting bosons after the frequency $\omega$ is quenched to zero. (Left) the solid lines indicates the analytical plot of the scaling solution $\lambda(t)^{-1}\, R(z/ \lambda(t)) \, \equiv x_0 \rho(x,t)$ with $z=x/x_0$,  given in Eq.~\eqref{eq:rscale} at various times $t$, with the parameter $\lambda(t)$ implicitly defined in Eq.~\eqref{eq:lambda}. The dashed black lines are obtained by numerically solving the time-dependent GPE [Eq.~\eqref{eq:TDGPE}] (detailed in~\ref{A:sim_free_exp}). (Right) The scaled collapsed plot of the average density profile in log-linear scale with the solid orange line plotting $R(y) \equiv \lambda(t) \, x_0 \rho(x,t)$ with $y=x/[x_0\lambda(t)]$, given in Eq.~\eqref{eq:TF_density}, and the markers obtained by numerically solving the time dependent GPE.
The following parameters are used in the simulations, $N = 5 \times 10^5$, $m=0.4$, $\omega=0.1$, $g=0.01$. } 
    \label{fig:i_dyn}
\end{figure}

In particular, the limiting behaviour of $\lambda(t)$ can be easily found as 
\begin{equation}
\label{eq:lam_lim}
    \lambda(t) \, =\begin{cases}
        \displaystyle
            1 + \frac{1}{2}(\omega t)^2  + O((\omega t)^4)&  \text{as}~~\omega t\to 0 \, ,\\[5mm] \displaystyle
        \sqrt{2} \, \omega t  - \frac{1}{2}\ln(\omega t) + O(1) & \text{as}~~ \omega t \to\infty, 
    \end{cases}
\end{equation}
Therefore, while at short times, the gas expands super-ballistically, eventually, it  expands ballistically at long times.

Once, $\lambda(t)$ is obtained by inverting Eq.~\eqref{eq:lambda}, the time-dependent density profile in the absence of resetting is given by Eq.~\eqref{eq:rscale}, with the scaling function $R(z)$ defined in Eq.~\eqref{eq:TF_density}.
Thus, we find that the expansion of the gas is self-similar, and the average density profile is an inverted parabola at all times. Note that $t(\lambda)$ in Eq.~\eqref{eq:lambda} describes the time at which the support of function $\rho(x,t)$ given in Eq.~\eqref{eq:rscale} reaches the points $ \pm x_0 \lambda$.

We compare the scaling solution of the time-dependent density profile in Eq.~\eqref{eq:rscale} with numerically evolved solution of the time-dependent GPE (details in \ref{A:sim_free_exp}) in Fig.~\ref{fig:i_dyn}, and find an excellent agreement. Having obtained the time-dependent density profile in the absence of resetting, we compute the density profile, edge statistics, and the full counting statistics in the NESS, in the presence of resetting in the next section [Sec.~\ref{s:observables}]. 

\section{NESS in the presence of stochastic resetting}
\label{s:observables}
In this section, we first study the density profile in the presence of resetting. Next, we use the density profile in the NESS to study the edge fluctuations and the full counting statistics.

\subsection{Steady state density profile}
\label{subsec:ssdp}

The steady state density profile of the gas in the presence of resetting is related to that in the absence of resetting via the renewal equation
[see~\ref{app:resetting}]
\begin{equation}
\label{eq:den_ren1}
    \rho_r(x) =  r\int_{0}^{\infty} \, d\tau \, e^{-r \tau } \rho(x,\tau) \, , 
\end{equation}
where $\rho(x,t)$ is given in Eq.~\eqref{eq:rscale}. It is evident from Eq.~\eqref{eq:rscale} that the steady state density profile $\rho_r(x)$ in Eq.~\eqref{eq:den_ren1} admits the scaling form
\begin{equation}
\label{eq:den_scal}
    \rho_r(x) = \frac{1}{x_0} \, G\left(\frac{x}{x_0}\right) \, ,
\end{equation}
with the scaling function $G(z)$ given by  
\begin{equation}
\label{eq:G_zt}
    G(z) =  r \int_{0}^{\infty} \, d\tau \,  e^{-r \tau}  \frac{1}{ \lambda(\tau)} \, R{\left( \frac{z}{ \lambda(\tau)}\right)} \, ,
\end{equation}
where $R(z)$ is defined in Eq.~\eqref{eq:TF_density} and $\lambda(\tau)$ is obtained by inverting Eq.~\eqref{eq:lambda}, i.e., the solution of the following equation
\begin{equation}
\label{eq:lambda-tau}
 \frac{1}{\sqrt{2}\, \omega }  \left[   \sqrt{\lambda(\lambda-1)} \, + \ln \left(\sqrt{\lambda}+\sqrt{\lambda -1}\right) \right] =\tau\, .
\end{equation}

It is convenient to make a change of the integration variable from $\tau$ to $\lambda(\tau)$ given in Eq.~\eqref{eq:lambda-tau}. This is equivalent to going from the random variable $\tau$ drawn from $p(\tau)=r\, e^{-r\tau}$ to another random variable $\lambda$ drawn for the PDF $h(\lambda)$. The PDF $h(\lambda)$ is related to $p(\tau)$ by $h(\lambda) = p(\tau(\lambda)\, |d\tau(\lambda)/d\lambda|$. Using Eqs.~\eqref{eq:beta1a} and \eqref{eq:lambda-tau}, after some algebra, we get
\begin{align}
    \label{eq:h_lam}
    h(\lambda) = r_0 \sqrt{\frac{\lambda}{\lambda-1} }  \, \left(\sqrt{\lambda } - \sqrt{\lambda-1} \right)^{r_0} \, e^{-r_0 \sqrt{\lambda(\lambda-1)}}    \quad \text{with} \quad r_0 = \frac{r}{\sqrt{2} \omega } \, .
\end{align}
The support of $h(\lambda)$ lies in $\lambda \in [1,\infty)$. The cumulative distribution of $h(\lambda)$  is given by
\begin{equation}
    F(\lambda)= \int_{\lambda}^\infty h(\lambda')\, d\lambda'= \left(\sqrt{\lambda } - \sqrt{\lambda-1} \right)^{r_0} \, e^{-r_0 \sqrt{\lambda(\lambda-1)}}.
    \label{eq:F}
\end{equation}
The normalization $\int_1^\infty h(\lambda)\, d\lambda = F(1)=1$ is easily checked.
Therefore, making a change of variable from $\tau$ to $\lambda$ in Eq.~\eqref{eq:G_zt} yields
\begin{equation}
    G(z) = \int_1^\infty d\lambda\, h(\lambda)\,\frac{1}{\lambda} R\left(\frac{z}{\lambda}\right)\,  ,
    \label{eq:Gzzz}
\end{equation}
with $h(\lambda)$ given in Eq.~\eqref{eq:h_lam}. Eq.~\eqref{eq:Gzzz}  encodes the effects of interplay between the attractive correlation generated by resetting and the repulsive interaction between the gas in the density profile $\rho_r(x) = x_0^{-1} G(x/x_0)$, given in  Eq.~\eqref{eq:den_scal}. The resetting rate $r$ appears only in the function $h(\lambda)$ given in Eq.~\eqref{eq:h_lam}, whereas the repulsive interaction strength $g$ appears through $x_0$ [Eq.~\eqref{eq:x0-2}] in the scaling function $R\bigl(x/(x_0 \lambda)\bigr)$.   
Inserting $R(z)$ from Eq.~\eqref{eq:TF_density} in  Eq.~\eqref{eq:Gzzz}, one explicitly gets 
\begin{equation}
    \label{eq:G_z}
    G(z) = \frac{3}{4}\int _{1}^{\infty} \, d\lambda \, h(\lambda)\frac{\left( \lambda^2 - z^2\right)}{ \lambda^3}  \, \Theta{\left( \lambda^2 - z^2 \right)} \, . 
\end{equation}

We plot $h(\lambda)$ in \fref{fig:lam}, and it has the following limiting behaviour
\begin{equation}
\label{eq:h(lambda)_lim}
    h(\lambda) \sim \begin{cases}
        \displaystyle 
        \frac{r_0}{ \sqrt{\lambda-1}} & \text{as }~\lambda\to 1,\\[5mm]
        \displaystyle
        r_0\left(\frac{ e }{4 \lambda}\right)^{r_0/2}\, e^{- r_0 \lambda} &\text{as }~ \,  \lambda\to\infty 
        \end{cases}
\end{equation}
\begin{figure}
    \centering
   \includegraphics[width=0.49\textwidth]{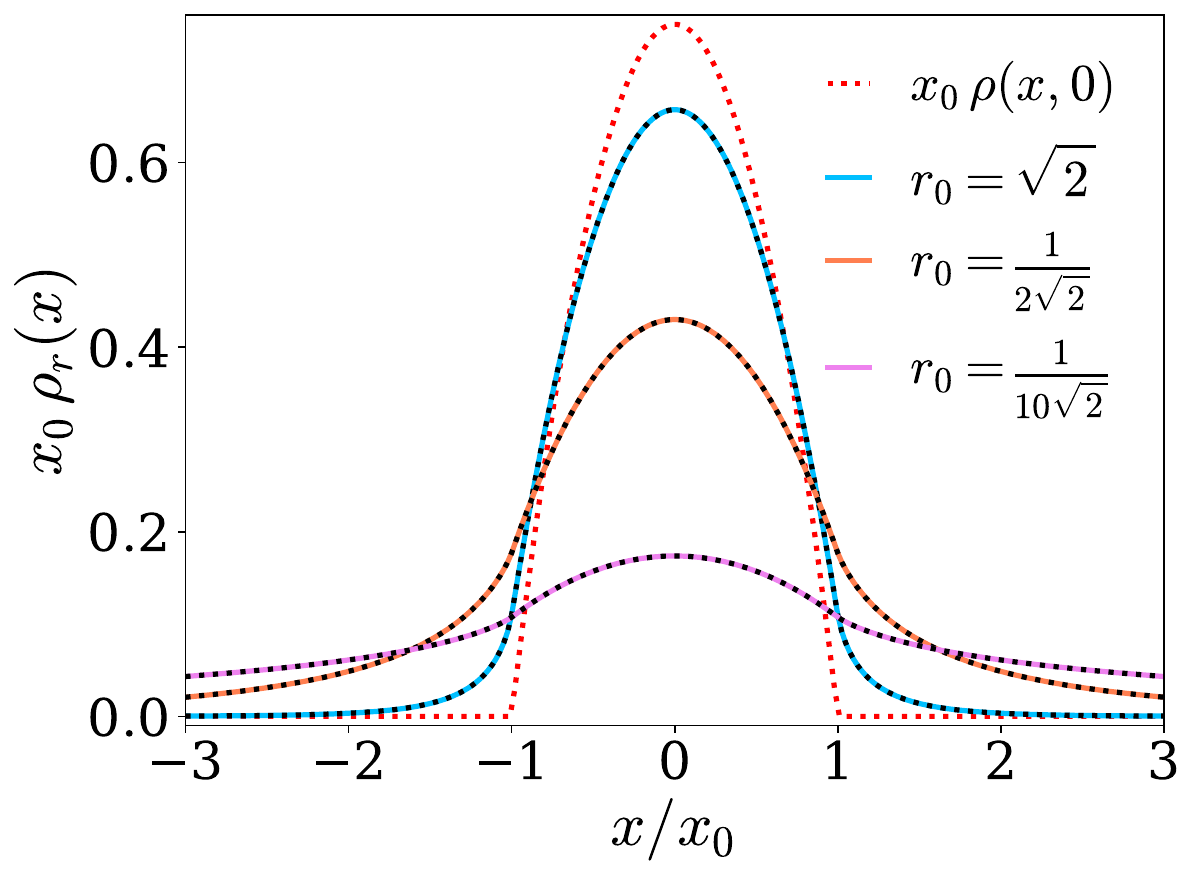}
\includegraphics[width=0.49\textwidth]{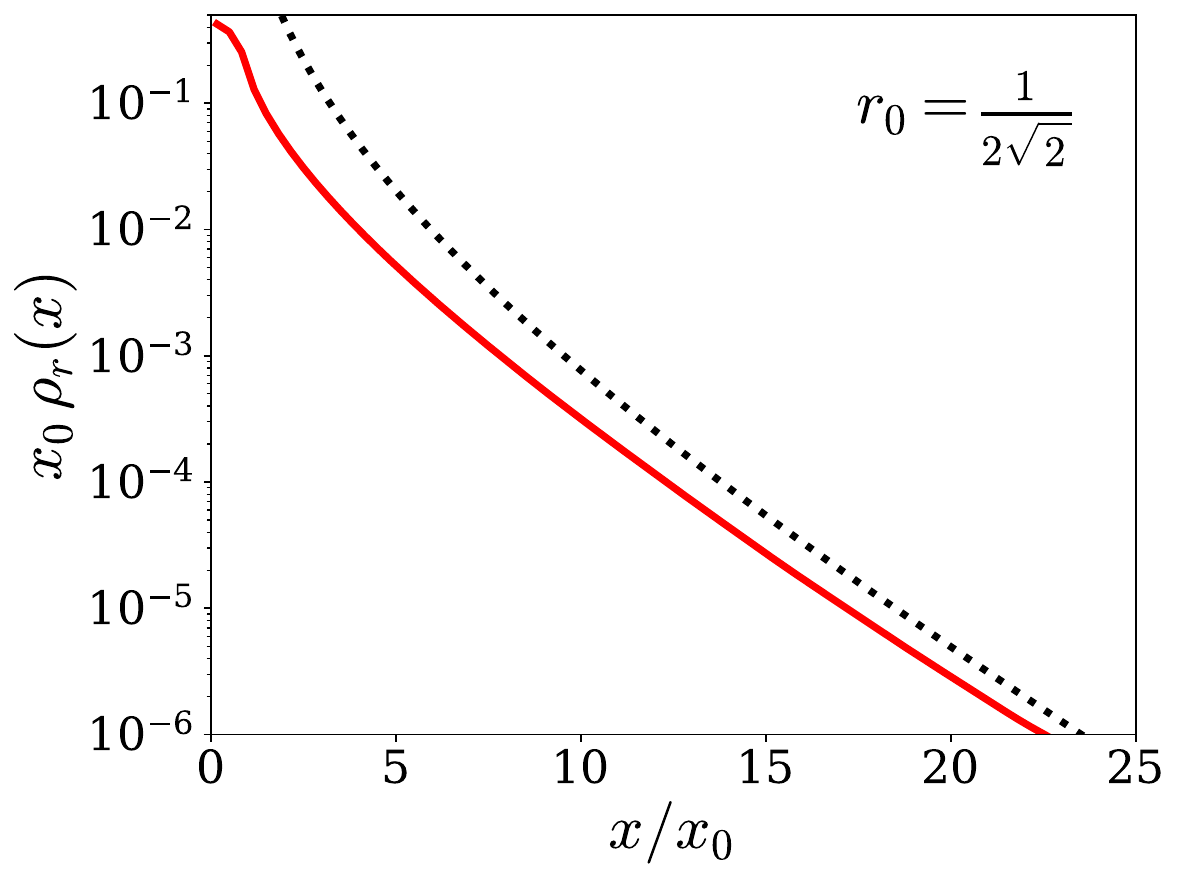}
    \caption{NESS condensate density of weakly interacting bosons subjected to stochastic resetting as a function $z=x/x_0$, in (left) linear-linear plot  and (right) log-linear plot for $x/x_0>0$.
    The solid colored lines are obtained by simulations detailed in \ref{A:sim_den_r}. (Left) The dashed black line is the plot of the analytical function $G(z) \equiv x_0 \rho_r(x)$ [Eq.~\eqref{eq:G(z)}] obtained by performing the numerical integration in Eq.~\eqref{eq:G(z)}.  For comparison we  also plot the initial density profile $R(z) \equiv x_0 \, \rho(x)$ [Eq.~\eqref{eq:TF_density}] in the dotted red line. (Right) The dashed black line plots the asymptotic form of $G(z)$ in Eq.~\eqref{eq:den_asym}. 
    The simulations parameters are $x_0 = 1$, $\omega =1$, and various $r$, recalling that $r_0 = r/(\sqrt{2}\omega)$, and the density profiles have been averaged over $10^5$ realizations (left) and $10^6$ realizations (right).} 
        \label{fig:den_r}
\end{figure}

The scaling function $G(z)$ in Eq.~\eqref{eq:G_z}  has two qualitative behaviours, given by 
\begin{equation}
\label{eq:G(z)}
    G(z) =  
    \begin{cases}\displaystyle
    \frac{3}{4}\int_{1}^{\infty} d\lambda \,  h(\lambda) \frac{1}{\lambda} \left( 1 - \frac{z^2}{ \lambda^2}\right) \quad \text{for } \, |z|<1 \, , \\[5mm] \displaystyle
    \frac{3}{4} \int_{|z|}^{\infty} \, d\lambda \, h(\lambda)\frac{1}{\lambda} \, \left( 1 - \frac{z^2}{ \lambda^2}\right) \quad \text{for} \, |z| > 1 \, .
    \end{cases}
\end{equation}
In the region $|z|<1$, $G(z)$ in Eq.~\eqref{eq:G(z)} retains its inverted parabolic shape, i.e., 
\begin{equation}
\label{eq:G_z_small}
    G(z) = \frac{3}{4} [a(r_0) - b(r_0) z^2 ] \, \quad \text{for} \quad |z| \leq 1 \, ,
\end{equation}
where
\begin{equation}
 a(r_0) =   \int_1^\infty \frac{h(\lambda)}{\lambda}\, d\lambda\quad\text{and}\quad b(r_0) = \int_1^\infty \frac{h(\lambda)}{\lambda^3}\, d\lambda \, ,
 \label{eq:abr0}
\end{equation}
with $h(\lambda)$ given in Eq.~\eqref{eq:h_lam}.

It is interesting to note that while the rescaled density profile given by the scaling function $G(z)$ in Eq.~\eqref{eq:G(z)} and its first derivative are both continuous at $z=\pm 1$, its second derivative is discontinuous. Indeed, from Eq.~\eqref{eq:G(z)}, it follows that 
\begin{equation}
    G''(z) =  \begin{cases}\displaystyle
        -\frac{3}{2} b(r_0)  &\quad \text{for } \, |z| < 1 \, \\[5mm]
        \displaystyle
       \frac{3 h(|z|)}{2 z^2} - \frac{3}{2} \int_{|z|}^\infty \frac{h(\lambda)}{\lambda^3}\, d\lambda   &\quad \text{for } \, |z|>1 \, 
    \end{cases}
    \label{eq:disc}
\end{equation}
where $b(r_0)$ is given in Eq.~\eqref{eq:abr0}. 
As $|z| \to 1^+$, $h(|z|)$ diverges as in Eq.~\eqref{eq:h(lambda)_lim}. Hence, as $|z|\to 1$, the second derivative $G''(z)$ remains finite for $|z| \to 1^-$, while it diverges as $|z|\to 1^+$. 
We recall that $|z|=1$ was the boundary of the initial Thomas-Fermi density profile, where the first derivative is discontinuous at $|z|=1$ . Thus, it is interesting to note that this edge singularity in the initial density profile is still remembered by the system at long times (in the NESS). However, the initial edge-singularity gets partially smoothened (the first derivative becomes continuous), but still with the second derivative discontinuous across $|z|=1$.

We next derive the asymptotic behavior of $G(z)$ for $|z| \gg 1$. To study this asymptotic behaviour, we make a change of variable $\beta = \lambda/z$, which reduces Eq.~\eqref{eq:G(z)} to 
\begin{equation}
    G(z) = \frac{3}{4 } \int_{1}^{\infty} \, \frac{d\beta}{\beta} \,  h{( z\beta)} \left(1- \frac{1}{\beta^2} \right) \, .
\end{equation}
In the large $z$ limit, we can use the large argument form of $h(z)$ from Eq.~\eqref{eq:h(lambda)_lim}, leading to 
\begin{equation}
\label{eq:G_z_asym}
    G(z) \simeq \frac{3 r_0 e^{r_0/2}}{ \, 2^{ 2 + r_0 } \, z^{r_0/2} } \, \int_{1}^{\infty} \, d\beta \, \beta^{-r_0/2  -3} \, (\beta^2-1) \, e^{- r_0 z \beta} \, .
\end{equation}
Making a change of variable $\beta = u+1$, one obtains 
\begin{equation}
    G(z) \simeq \frac{3r_0 \, e^{-r_0 (z-1/2) } }{2^{r_0+2} z^{r_0/2}} \int_{0}^{\infty} \, du \frac{u(u+2)}{(u+1)^{3 +r_0/2}} \, e^{-r_0 z \,u}  .
    \label{eq:G(z)-wx}
\end{equation}
It is evident from the exponential tail $e^{-r_0 z u}$ in the integrand that the integral for large $z$, is dominated by the small $u$ behavior of the prefactor of the exponential, i.e., $u (u+2)(u+1)^{-3-r_0/2} = 2u -(5+r_0)u^2 + O(u^3)$. Using this in Eq.~\eqref{eq:G(z)-wx}, and performing the integrals term by term, to the leading order,  we get 
\begin{equation}
\label{eq:den_asym}
    G(z) \simeq   \, B \,  \,z^{-2 - r_0/2} \, e^{-r_0 z} \quad \text{for large $z$} \, ,~~\text{where}~~ B=\frac{3 \, e^{r_0/2} }{  \,  2^{r_0+1} \, r_0 } ~~\text{and}~~r_0 = \frac{r}{\sqrt{2} \, \omega} \, .
\end{equation}
Thus, the sharp support of the Thomas-Fermi initial density profile in Eq.~\eqref{eq:TF_density} is smoothened out to exponential decay by the resetting protocol.

In Fig.~\ref{fig:den_r}, we  compare the density profile given in Eq.~\eqref{eq:G(z)}, with simulation and find excellent agreement. 
We also compare the asymptotic form of the density profile given in Eq.~\eqref{eq:den_asym}  with simulation data in Fig.~\ref{fig:den_r} and find good agreement. To summarize, so far, we find that
\begin{enumerate}
\item The steady-state density profile $\rho_r(x)$ in the presence of resetting has the scaling form $\rho_r(x) = x_0^{-1}\, \, G(x/x_0)$, where $x_0 \propto (gN)^{1/3}$ sets the length scale, with $g$ being the interaction strength. 
    \item The scaling function $G(z)$ that has two different functional forms, one for $|z|<1$, and another for $|z|>1$.
    \item For $|z|\le 1$, the scaling function $G(z) \propto  a - b \,z^2$ is always an inverted parabola [Eq.~\eqref{eq:G_z_small}].
    \item For $z \gg 1$ the scaling function $G(z) \sim  e^{-r_0 z}$, has an exponentially decaying tail,  given in Eq.~\eqref{eq:den_asym}. This implies 
    $\rho_r(x) \sim e^{-x/\xi}$ with $\xi = x_0/r_0  \propto g^{1/3}/r$, capturing the interplay between the interaction strength $g$ and the resetting rate $r$.
\end{enumerate}

\subsection{Edge fluctuations in the steady state}
\label{s:edge}

Let $x_{\max}$ be the position of the rightmost particle of a system of $N$ weakly interacting bosons. In the absence of resetting, the gas undergoes  free expansion and we can write $x_{\max}(t) = \overline{x_{\max}}(t) + \delta x_{\max}$, where $\overline{x_{\max}}(t)$ represents the quantum expectation value of $x_{\max}(t)$ and $\delta x_{\max}$ denotes quantum fluctuations.
As we observed in \sref{s:no_resetting}, the density profile during the free expansion retains the inverted-parabola shape, with the support expanding as $x_0 \lambda(t)$, where $x_0$ and $\lambda(t)$ are given in Eqs.~\eqref{eq:x0-2} and \eqref{eq:lambda} respectively. Therefore,  $\overline{x_{\max}}(t)=x_0 \lambda(t)$, i.e., the quantum expectation value of $x_{\max}$ coincides with the right edge of the macroscopic density profile. In the presence of resetting, the time $\tau$ since the last resetting event is a random variable drawn from $p(\tau)=r\, e^{-r\tau}$. Therefore, $\overline{x_{\max}}(\tau)=x_0 \lambda(\tau)$ is also still a random variable.

To determine the quantum fluctuations $\delta x_{\max}$, one needs to solve the full quantum many-body dynamics in Eqs.~\eqref{eq:H(w)_field}--\eqref{eq:sq_ham}. This is very hard. 
However, in a recent series of works it was found
for both  classical~\cite{BLM2023,BLM2024,BLM2024_2, SM2024, MMS2025, dBM2026, DS2025, BMS2025, Olsen2026, BCK2025,   deMauro_2026, galla2026, demauro2026effects} and quantum~\cite{MCP2022, KMS2025, DIG2025, MKM2025} many-body systems with DEC that $x_{\max}$ in the absence of stochastic resetting has a dominant deterministic part and a sub-dominant fluctuating part. The sub-dominant fluctuating part $\delta x_{\max}$ does not contribute to the PDF of $x_{\max}$ in the presence of resetting. One can perhaps  appeal to the example of the non-interacting gas of $N$ fermions in a harmonic trap at zero temperature~\cite{Dean_2019}. Although there is no direct interactions between the fermions, the Pauli exclusion principle effectively makes the system correlated where the JPDF of the position of the fermions is non-factorizable, akin to the JPDF of the eigenvalues of Gaussian unitary ensemble of random matrix theory~\cite{mehta2004random, forrester2010log}. In this case, the density profile is given by the Wigner semi-circle law, where the support scales as $\sqrt{N}$. The fluctuation of the edge  particles around the support is described  by the well-known Tracy-Widom distribution~\cite{TW1993} at the scale of $N^{-1/6}$. Therefore, this fluctuation is subdominant compared to scale $\sqrt{N}$ of the support. This was also verified recently in the context of the resetting Dyson Brownian gas~\cite{BMS2025_2}. For the weakly interacting, freely expanding Bose gas considered in this work, the right edge $\overline{x_{\max}}(t)=x_0 \lambda(t)$ scales as $N^{1/3}$. The GPE framework is ill-suited for computing the quantum fluctuations $\delta x_{\max}$ around the edge, since it is essentially an effective mean-field theory. To compute the full quantum fluctuations of observable, one needs to study the full microscopic description in Eqs.~\eqref{eq:H(w)_field}--\eqref{eq:sq_ham}, which is very hard. However, based on the examples cited above, it is reasonable to assume that fluctuations around the edge (of order $N^{1/3}$) are subdominant in the large $N$ limit.

\begin{figure}
    \centering
    \includegraphics[width=.5\linewidth]{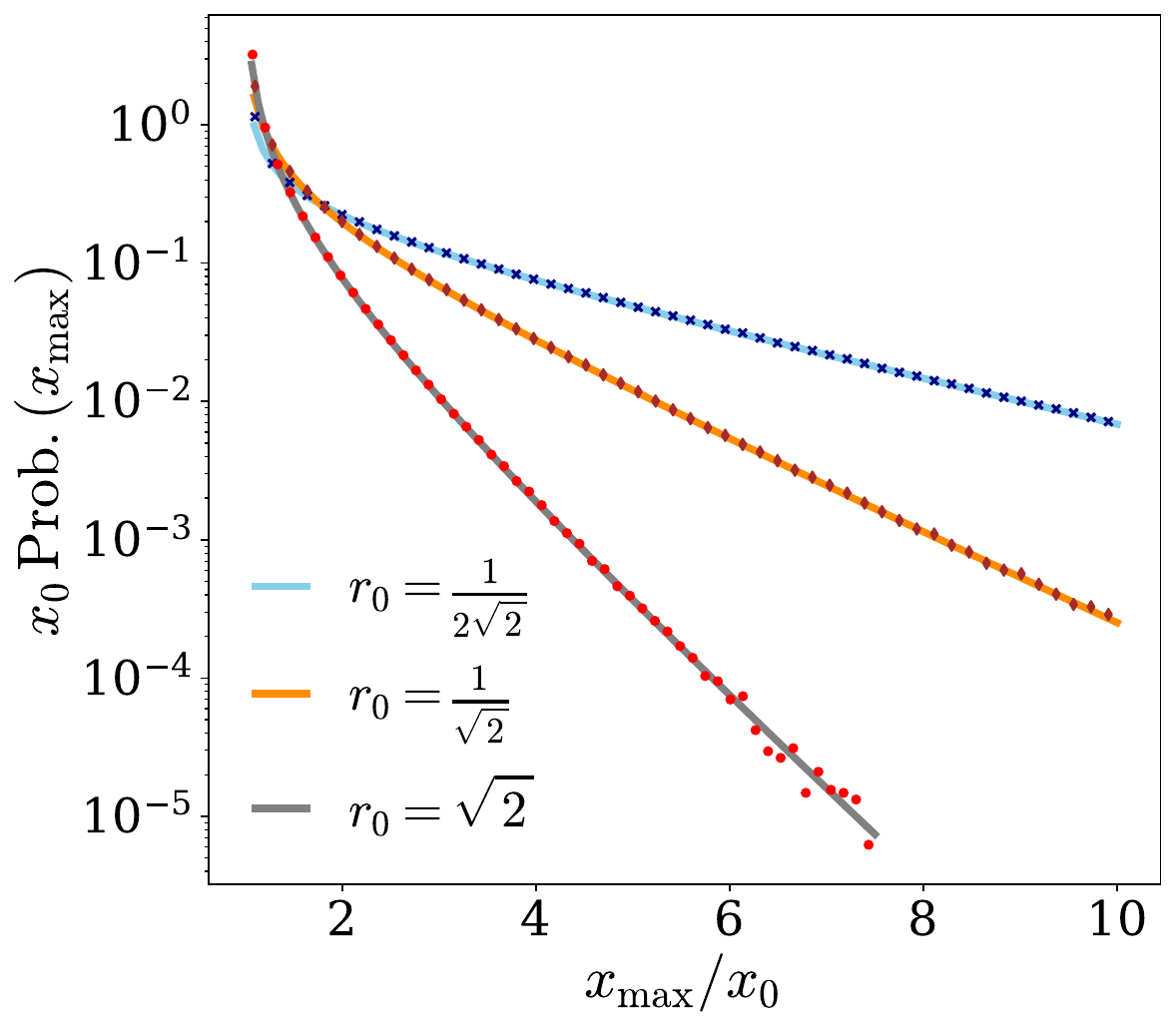}
    \caption{The edge support distribution of the condensate density in the NESS, as a function of $z=x_{\max}/x_0$. The solid line is the analytical function $ h(z) \equiv x_0 \mathrm{Prob}.(x_{\max})$ given in Eq.~\eqref{eq:h_lam}. The markers are obtained by picking $t$ from the random distribution $r e^{-r t}$, and finding $\lambda(t)$ by numerically solving Eq.~\eqref{eq:lambda} with $\omega$ set to unity. This procedure is then repeated over $10^7$ realizations, and the data is binned, averaged and plotted.} 
    \label{fig:edge}
\end{figure}

Although, the density profile $\rho(x,t)$ under the free expansion in the GPE framework, does not have any quantum fluctuations around  $\overline{x_{\max}}(\tau)= x_0\lambda(\tau)$, when subjected to resetting, the time $\tau$ from the last reset event becomes a random variable drawn from the PDF $p(\tau)= r\, e^{-r \tau}$. Therefore, resetting makes $\overline{x_{\max}}(\tau)$  also a random variable that varies from one realization to another. Consequently, the distribution of the rightmost particle $x_{\max}$ is given by 
\begin{equation}
    \mathrm{Prob}.(x_{\max}) = r \int_{0}^{\infty} d\tau \, e^{-r \tau } \, \delta[ x_{\max} - \overline{x_{\max}}(\tau) ] = \int_{1}^{\infty} d\lambda  \, h(\lambda) \, \delta[ x_{\max} - x_0 \lambda] \,,
    \label{eq:Pxmax}
\end{equation}
where we have made a change of variable from $\tau$ to $\lambda$ using Eq.~\eqref{eq:lambda}.
Performing the integration over $\lambda$ in Eq.~\eqref{eq:Pxmax} yields
\begin{equation}
    \mathrm{Prob}.(x_{\max}) = \frac{1}{x_0} \, h\left(\frac{x_{\max}}{x_0} \right) \, ,
\end{equation}
where we recall that $h(\lambda)$ is defined in Eq.~\eqref{eq:h_lam} and is plotted in Fig.~\ref{fig:lam}. Since the support of the right edge is $x_0$ at $t=0$ and the gas is repulsive, $x_{\max} > x_0$. This is, in fact, also needed, as the support of the function $h(z)$ is  $z\in [1,\infty)$. The asymptotic behavior of $\mathrm{Prob}.(x_{\max})$ as $x_{\max}\to x_0$ and $x_{\max}\to\infty$ follows from the asymptotic behaviour of $h(\lambda)$ given in Eq.~\eqref{eq:h(lambda)_lim}. In Fig.~\ref{fig:edge} we plot the distribution of $x_{\max}$ and compare it with simulations.

\subsection{Full counting statistics in the steady state}
\label{s:FCS}
\begin{figure}
    \centering
    \includegraphics[width=0.49\textwidth]{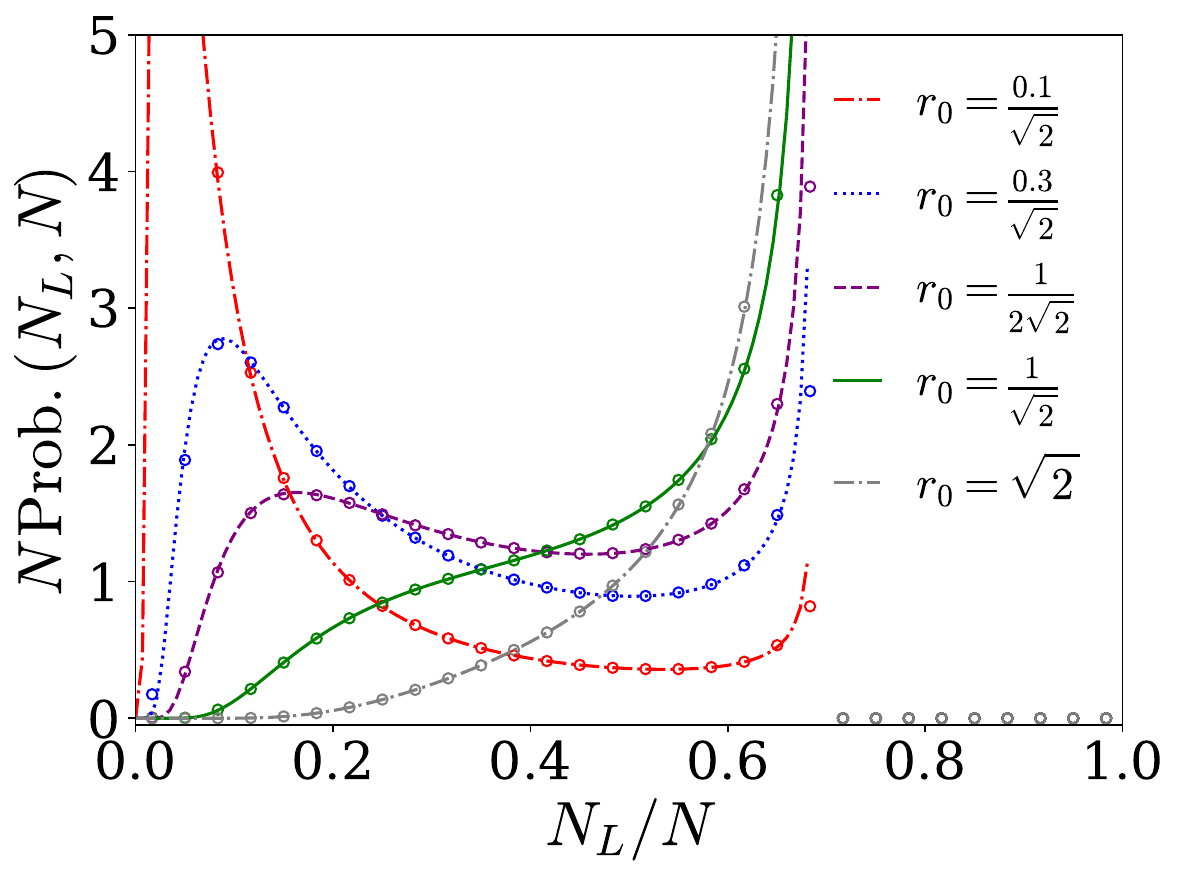}~   \includegraphics[width=0.49\textwidth]{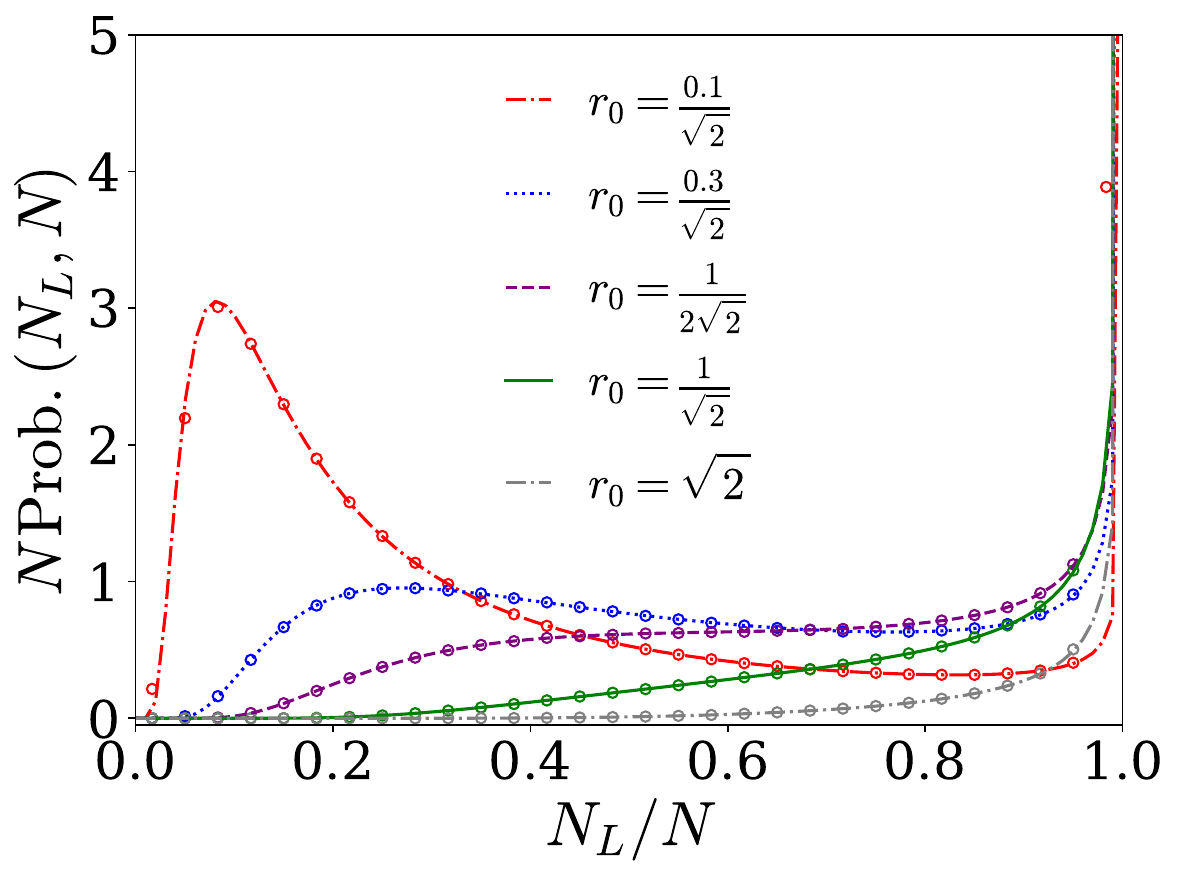}    
    \caption{NESS PDF of the fraction of the condensate bosons $\kappa=N_L/N$ within the domain $[-L,L]$ for $l =0.5 $ (left) and $l=1.5$ (right), where $l= L/x_0$ [$x_0$ defined in Eq.~\eqref{eq:x0-2}]. The dashed lines are plots of the regular part of the function $ N \, \mathrm{Prob.}(N_L, N) \equiv H_{\mathrm{reg}}(\kappa)$ given in Eq.~\eqref{eq:Hkappa}, while the markers are obtained via direct numerical simulations described in section \ref{A:sim_den_r}. The simulations are performed with parameters $x_0=1$, $\omega =1$, $r_0 = r/(\sqrt{2}\,\omega)$ and averaged over $10^{7}$ realizations.}
    \label{fig:FCS}
\end{figure}

Finally, we study the distribution of the number of particles $N_L$ confined within a volume $[-L,L]$, known as the full counting statistics (FCS), in the NESS. 
As in \sref{s:edge} on the statistics of the edge, we obtain the distribution of $N_L$ by considering the fluctuations arising solely due to the resetting protocol, neglecting the subdominant microscopic quantum fluctuations, i.e., $N_L=\overline{N}_L + \delta N_L$, where $\overline{N}_L$ represents the quantum expectation value of $N_L$, and we neglect $\delta N_L$. Therefore, 
the number of particles confined within $[-L,L]$ at a given time $\tau$ since the last resetting event is 
\begin{equation}
  \overline{N}_L (\tau) = N\int_{-L}^{L}  \, \rho(x,\tau)\,  dx,
  \label{eq:NL}
\end{equation}
where $\rho(x,\tau)$ is given in Eq.~\eqref{eq:rscale} along with Eq.~\eqref{eq:TF_density}. Since $\tau$ is a random variable drawn from $p(\tau)=r e^{-r\tau}$, the quantity $\overline{N}_L(\tau)$ in Eq.~\eqref{eq:NL} is still a random variable. 
Therefore, the  FCS in the NESS within the GPE framework is given by 
\begin{equation}
    \mathrm{Prob}.(N_L, N) = r\int_{0}^{\infty} \, e^{-r \tau} \delta [ N_L - \overline{N}_L(\tau)] \, d\tau= \int_{1}^{\infty} d\lambda \, h(\lambda) \, \delta[ N_L - \overline{N}_L(\tau(\lambda))] \, ,
\end{equation} 
where $\tau(\lambda)$ and $h(\lambda)$ are given in  Eqs.~\eqref{eq:lambda} and \eqref{eq:h_lam} respectively.  It is evident that $\mathrm{Prob}.(N_L, N)$ has the scaling form 
\begin{equation}
\label{eq:FCS3}
    \mathrm{Prob.}(N_L, N) = \frac{1}{N} \, H{\left( \frac{N_L}{N}\right)} \, ,
\end{equation}
where the scaling function $H(\kappa)$ is given by
\begin{equation}
    H(\kappa) = \int_{1}^{\infty} d\lambda \, h(\lambda) \, \delta[\kappa - \kappa(\lambda)] \, ,
    \label{eq:Hkappa_0}
\end{equation}
with $\kappa(\lambda) =  N_L(\lambda)/N$ being the fraction of particles in $[-L,L]$. Skipping details [see \ref{app:FCS}], we find
\begin{equation}
    H(\kappa) = \Theta(l-1)\, \bigl[1-F(l)\bigr]\, \delta(\kappa-1)  + \frac{ h(\lambda_*)\,l\,\cos\theta}{6 \, \sqrt{1-\kappa^2}\,\sin^2\theta} \, \Theta(\kappa)\, \Theta(\kappa_{\max}-\kappa),
    \label{eq:Hkappa}
\end{equation}
where $h(\lambda)$ and $F(l)$ are  given in Eqs.~\eqref{eq:h_lam} and \eqref{eq:F}, respectively, and 
\begin{equation}
 l=L/x_0, \quad  \lambda_*(\kappa) = \frac{l}{2\sin\theta}\, , \quad\theta= \frac{1}{3}\arcsin(\kappa),  
   \label{eq:lambdastar}
\end{equation}
and the upper support
\begin{equation}
    \kappa_{\max}= \begin{cases}\displaystyle
        \frac{1}{2}(3l-l^3) &\quad\text{for}\quad l<1\, , \\[3mm]
        1  &\quad\text{for}\quad l \ge 1\, .
    \end{cases}
    \label{eq:kappamax}
\end{equation}
Thus, Eqs.~\eqref{eq:Hkappa}--\eqref{eq:kappamax} determine the FCS completely. Note that for $l>1$, the scaling function $H(\kappa)$ in Eq.~\eqref{eq:Hkappa} is sum of a singular part $H_\text{sing}(\kappa)$ containing a Dirac-delta function and a regular part $H_\text{reg}(\kappa)$.
In Fig.~\ref{fig:FCS}, we plot the distribution of FCS with regular part of the scaling function $H_\text{reg}(\kappa)$ and compare it with simulated data, and find a very good agreement.

Near the lower support $\kappa\to 0$, we have [see \ref{app:FCS}]
\begin{equation}
    H(\kappa) \sim   \kappa^{r_0/2-2}\, \exp\left(-\frac{3r_0l}{2\kappa}\right),
\end{equation}
which has an essential singularity at $\kappa=0$. 
On the other hand, near the upper support $\kappa\to \kappa_{\max}$ , the regular part of $H(\kappa)$ behaves as [see \ref{app:FCS}]
\begin{equation}
    H_\text{reg}(\kappa) \sim  \begin{cases}\displaystyle
    \frac{1}{\sqrt{1-\kappa}} &\quad\text{for}~~ l>1,\\[3mm]
    \displaystyle
    \frac{1}{(1-\kappa)^{3/4}} &\quad\text{for}~~ l=1,\\[3mm]
    \displaystyle
    \frac{1}{\sqrt{\kappa_{\max}-\kappa}} &\quad\text{for}~~ l<1,
    \end{cases}
\end{equation}
where $\kappa_{\max}$ is given in Eq.~\eqref{eq:kappamax}.

\section{Conclusions and Outlook}
\label{sec:conc}

In this work, we have investigated stochastic resetting in an interacting quantum many-body system, focusing on a weakly repulsive Bose--Einstein condensate described within the Gross-Pitaevskii framework. The central motivation was to understand how  attractive correlations generated dynamically by a common fluctuating environment competes with intrinsic repulsive interactions between particles. While simultaneous resetting can generate strong, effectively attractive dynamically emergent correlations (DEC) even in an otherwise noninteracting system, the bosons considered here are inherently repulsive. Our model therefore provides a rich, yet analytically tractable setting in which these two distinct sources of correlations are simultaneously present in a nonequilibrium steady state (NESS).

We prepare our system in the Thomas-Fermi initial state. This choice is motivated both physically and analytically. Thomas-Fermi state represents the natural
stationary state of the harmonically trapped condensate prior to its
release, making it experimentally amenable. Moreover, this particular initial condition has an important advantage, once  the trap is switched off. The density evolves
self-similarly, retaining its inverted-parabolic form while its support is
controlled by a single time-dependent scale factor $\lambda(t)$. This
reduces the GPE dynamics to an effective equation for
$\lambda(t)$ and is the key property that allows us to obtain the subsequent
resetting problem analytically. For a generic initial state, such a closed
self-similar evolution need not hold.

In the absence of resetting, 
the gas expands from the Thomas-Fermi initial state according to the Gross-Pitaevskii equation. Interestingly, the time-dependent density profile during the expansion admits a self-similar form that is dictated by a single scale factor $x_0\,\lambda(t)$, where $\lambda(t) \ge 1$ is an increasing function of $t$ and is completely independent of the interaction strength $g$.
The interaction enters only through the Thomas-Fermi length
$x_0\propto (gN)^{1/3}$, which fixes the initial spatial scale of the cloud.
This separation between an interaction-dependent length
scale $x_0$ and an interaction-independent dynamical scaling factor $\lambda(t)$ is the key
simplification that makes the problem analytically tractable, despite the
presence of interactions. 
We have shown that stochastic resetting drives the freely expanding condensate into a unique NESS whose spatial structure retains clear signatures of both the interactions and the resetting protocol. The stationary density assumes the scaling form $\rho_r(x)=x_0^{-1}G(x/x_0)$, where the Thomas-Fermi length $x_0$ encodes the intrinsic repulsive interaction. Remarkably, inside the original Thomas-Fermi support $[-x_0,x_0]$, the density retains an inverted-parabolic form, whereas outside this region resetting removes the sharp edge of the initial condensate and generates an exponentially decaying tail. The interplay between the direct repulsive interaction between the bosons and the attractive correlation generated by resetting shows up in the length scale of the exponential decay of the tail of the NESS density profile $\rho_r(x) \sim e^{-x/\xi}$ with $\xi = x_0/r_0  \propto g^{1/3}/r$.

Thus, stochastic resetting qualitatively reshapes the stationary state, producing a characteristic core--tail structure, where the  interplay between interactions and resetting can be clearly seen and studied analytically. Interestingly, the edge singularity in the initial density profile is still remembered by the system at long times. However, the initial edge-singularity gets partially smoothened, i.e.,  the first derivative $G'(z)$ becomes continuous, but still with the second derivative $G''(z)$ discontinuous across $|z|=1$.

Beyond the average density, resetting produces strong realization-to-realization fluctuations of several other interesting observables. In particular, the position of the rightmost (or leftmost) particle of the repulsive quantum bosonic gas under stochastic resetting is dictated by the condensate edge evolved from the last resetting event. Therefore, the condensate edge itself is a stochastic quantity whose stationary distribution is determined by the distribution $h(\lambda)$. We have also obtained the full counting statistics of the fraction of particles $\kappa=N_L/N$  contained in a finite interval $[-L,L]$. The PDF $H(\kappa)$ has an essential singularity near its lower support ($\kappa\to 0$) and divergent behavior near its upper support ($\kappa\to \kappa_{\max} \le 1$). For intervals larger than the initial Thomas-Fermi cloud, $H(\kappa)$ additionally acquires a Dirac delta contribution at $\kappa=1$. These results demonstrate that fluctuations generated by the common resetting environment survive at the macroscopic scale and leave pronounced signatures in both edge and number statistics. An appealing feature of the problem is that the same function $h(\lambda)$ underlies the stationary density, edge statistics, and full counting statistics, providing a unified description of several seemingly different NESS observables.

A particularly promising direction is the fermionic counterpart of the present problem~\cite{fermion_future}. One may consider a harmonically trapped Fermi gas subjected to an analogous global resetting protocol, with the system repeatedly reset to its initial trapped many-body state. Even in the absence of direct interactions, fermions possess strong intrinsic correlations arising purely from the Pauli principle, making such a setup fundamentally different from both an ideal classical gas and the weakly repulsive Bose condensate studied here. The competition between these fermionic repulsive correlations and the attractive correlations generated by the common resetting environment could therefore provide another natural setting for investigating DEC in genuinely correlated quantum systems. Moreover, trapped fermions possess well-understood density, edge, and counting statistics---including universal edge fluctuations in the ground state~\cite{Dean_2019,TW1993}---raising the possibility of determining analytically how stochastic resetting modifies these structures in the resulting NESS. A comparison between bosonic and fermionic gases under otherwise analogous resetting protocols could help disentangle the respective roles of interactions, quantum statistics, and common environmental fluctuations in shaping correlated nonequilibrium states. Other natural extensions include going beyond the mean-field approximation to incorporate quantum fluctuations~\cite{MC2003}, and exploring stronger interactions where the Gross-Pitaevskii description breaks down~\cite{Petrov_2000}. It would also be interesting to investigate more general resetting protocols, including non-Poissonian resetting~\cite{dBM2026,BDC2016} and partial resets~\cite{GN2016,MCG2024,deMauro_2026,AMG2025}, which may lead to rich nonequilibrium behavior.

The ingredients of our proposed protocol involve standard techniques in cold-atom setups. Specifically, it uses the same basic procedure, the time-of-flight (TOF) technique~\cite{AEM1995, DMA1995, abs_img_1}, which involves the release of atoms by switching off the trap and the subsequent free expansion of the gas. 
Moreover, cold-atom systems offer an exceptionally controlled environment, 
where trapping potentials~\cite{Rohringer2015} and interaction strengths~\cite{CFH1998,IAS1998} can be controlled with high precision. Needless to mention, state-of-the-art absorption imaging~\cite{abs_img_1,abs_img_2,abs_img_3} and particle-resolved~\cite{quantum_microscopy_1,quantum_microscopy_2,quantum_microscopy_3} measurement techniques are possible in experiments. 
Measurements of the stationary density---in particular its resetting-induced tails and the second order discontinuity---together with edge statistics and particle-number fluctuations would provide direct probes of the predictions presented here. Measurements need not, however, be restricted to density observables. The condensate wavefunction also contains the phase $S(x,t)/\hbar$ defined in Eq.~\eqref{eq:cov}, whose values and spatial gradients are experimentally accessible. Phases can be probed through matter-wave interference~\cite{Andrews1997,Hofferberth2008,Berrada2013}, while the phase gradient, which determines the superfluid velocity through $v=(1/m)\partial_x S$, can be accessed using velocity-sensitive techniques such as Bragg spectroscopy~\cite{bragg1,Muniz2006}. More broadly, the present work establishes interacting ultracold gases as a useful platform for studying how stochastic resetting can impact correlations in quantum many-body systems and opens the way toward using controlled environmental fluctuations as a tool for engineering nonequilibrium correlated states.

\section*{Acknowledgments}
MK acknowledges the support of the Department of Atomic Energy, Government of India, under project
no. RTI4001. SNM and SS acknowledge the support from the Science and Engineering Research Board
(SERB, Government of India), under the VAJRA faculty scheme (No. VJR/2017/000110). SNM acknowledges support from ANR Grant No. ANR-23-CE30-0020-01 EDIPS. This research was supported in part by grant NSF PHY-2309135 to the Kavli Institute for Theoretical Physics (KITP). 
MK thanks the hospitality of Laboratoire de Physique Théorique et Modèles Statistiques (LPTMS), University Paris-Saclay and Collège de France, PSL Research University where a major part of the work took place.


\appendix

\section{Numerical Simulations}
\subsection{Ground state of the GPE}
\label{A:sim_gs}
In order to obtain the ground state wavefunction of the GPE in Eq.~\eqref{eq:GPE}, we start with time-dependent GPE in Eq.~\eqref{eq:TDGPE} with scaled spatial coordinate $y = x/x_0$ [$x_0$ defined in Eq.~\eqref{eq:x0-2}].
Setting $\hbar =1$, we write Eq.~\eqref{eq:TDGPE} in imaginary time,  $  t = -i \tau$, evolve the associated equation
\begin{equation}
\label{eq:IM_t}
    \partial_\tau \phi(y,\tau) = \left( \frac{1}{2m x_0^2} \frac{\partial^2}{\partial y^2} -  \frac{1}{2}m \omega^2 x_0^2 y^2 - \frac{N g}{x_0} |\phi(y,\tau)|^2 \right) \, \phi(y,\tau) \, 
\end{equation}
over a long time. This type of imaginary time propagation is quite standard for finding the ground state of the GPE~\cite{CST_2000}. In this case we use an initial wavefunction 
\begin{equation}
    \phi(y,0)= \frac{1}{(4 \, \pi)^{1/4}} \, \exp(- y^2/8) \, ,
\end{equation}
and numerically evolve this under the imaginary time GPE given in Eq.~\eqref{eq:IM_t}. This evolution is done by discretizing space with $\Delta y = 4 \times 10^{-3}$ and using an explicit Euler method with the time step $\Delta \tau = 1 \times  10^{-3}$, with the wavefunction properly normalized after each time step. The parameters used in this simulation are
$N = 5 \times 10^5$, $m=0.4$, $\omega=0.1$, $g=0.01$. The dashed black plot in Fig.~\ref{fig:gs} are obtained using this procedure, and the plot corresponds to the density $\rho(z) = |\phi(z,\tau)|^2$ at imaginary time $\tau=8$.

\subsection{Evolution of the wavefunction after the frequency quench}
\label{A:sim_free_exp}
We set $\hbar =1$, and write the post frequency [i.e., $\omega = 0$] GPE [Eq.~\eqref{eq:TDGPE1}], in terms of $y = x/x_0$ 
\begin{equation}
    i \frac{\partial}{\partial t} \phi(y,t) =  \left( - \frac{1}{2m x_0^2 } \frac{\partial^2}{\partial y^2} + \frac{Ng}{x_0}  |\phi(y,t)|^2 \, \right) \phi(y,t) \, ,
\end{equation}
where $x_0$ defined in Eq.~\eqref{eq:x0-2}.
To evolve this equation, we use the time-splitting spectral method, where we evolve the linear kinetic energy term, and the non-linear density term separately~\cite{WDP2003}. We introduce two equations 
\begin{equation}
    i \frac{\partial}{\partial t} \phi_1(y,t) = \frac{N g}{x_0} |\phi_1(y,t)|^2 \phi_1(y,t) \quad \text{and} \quad
    i \frac{\partial}{\partial t} \phi_2(y,t) = - \frac{1}{2 m x_0^2} \frac{\partial^2}{\partial y^2} \phi_2(y,t) \, , 
\end{equation}
and discretize time in units of $\Delta t$. In the first step we evolve $\phi_1(y,t)$ over half a step $\Delta t/2$ giving 
\begin{equation}
    \phi_1^*(y,t) = \exp\left( -\frac{iN g}{x_0} |\phi_1(y,t)|^2 \, (\Delta t/2)  \right) \phi_1(y,t) \, .
\end{equation}
In the second step we evolve with the kinetic term over a full time step $\Delta t$ as 
\begin{equation}
    \phi^*_2(y,t) = \exp \left(\frac{i \Delta t}{2 m x_0^2} \frac{\partial^2}{\partial y^2} \right) \, \phi_1^*(y,t) \, .
\end{equation}
Since the kinetic energy operator is diagonal in the momentum space, this procedure is efficiently done in the Fourier space.
Finally, we perform another half step in time with the non-linear term
\begin{equation}
    \phi(y\, ,t + \Delta t) = \exp\left( -\frac{iN g}{x_0} |\phi^*_2(y,t)|^2 \, (\Delta t/2)  \right) \phi^*_2(y,t) \, .
\end{equation}
This type of time stepping is called Strang splitting~\cite{WDP2003}. For our evaluation we have evolved the non-linear term in real space, while the derivative term has been evaluated in Fourier space.
This method has been implemented with $\Delta y = 4 \times 10^{-3}$ and $\Delta t = 1 \times 10^{-4}$ full Strang-time step.

\subsection{Simulation of density under resetting}
\label{A:sim_den_r}
To obtain the density profile of the NESS of the interacting Bose gas under resetting, first we pick $\tau$ from the random distribution $re^{-\ r\tau}$. The $\tau$ here plays the role of the time since the last reset. Now we find the corresponding $\lambda(\tau)$ by solving Eq.~\eqref{eq:lambda} numerically. Next, we obtain the corresponding density profile $\rho(x,t)$ using Eq.~\eqref{eq:rscale}. We repeat this procedure over $10^5 - 10^6$ realizations and average over them.

In order to obtain the distribution of the FCS, we follow the above procedure, and find the fraction of particles for each realization. The procedure is repeated over $10^7$ realizations, and the data is binned and plotted in Fig.~\ref{fig:FCS}.

\section{Density profile in the presence of stochastic resetting}
\label{app:resetting}

In this appendix, we recap the notion of quantum resetting and arrive at the renewal equation for the density profile. 
For any quantum many-body system, the key object is the density matrix $\hat\varrho(t)=\ket{\Psi(t)}\bra{\Psi(t)}$, which, in the absence of stochastic resetting, evolves from the initial density matrix   $\hat\varrho_0=|\Psi_0\rangle\langle\Psi_0|$ by the quantum unitary dynamics
\begin{equation}
    \hat\varrho(t) = e^{-(i/\hbar) \mathcal{H} t} \,\hat\varrho_0 \,e^{(i/\hbar) \mathcal{H} t}\, .
    \label{eq:formal}
\end{equation}
Stochastic resetting intermittently interrupts the unitary quantum dynamics at a constant rate $r$, resetting the density matrix to $\hat\varrho_0$, after which the quantum unitary evolution [Eq.~\eqref{eq:formal}] resumes from this initial state. One can show that the density matrix $\hat\varrho_r (t)$ under the stochastic resetting evolves as~\cite{MSM2018} 
\begin{equation}
    \hat\varrho_r (t) = e^{-r t} \, \hat\varrho(t) + r \int_0^{t} d\tau e^{-r \tau} \hat\varrho(\tau)\, ,
    \label{eq:rho_ren}
\end{equation}
where the subscript $r$ stands for `resetting'. Given the density matrix $\hat{\varrho}_r (t)$, the quantum JPDF of the positions of $N$ particles is given by the matrix element 
\begin{equation}
\label{eq:Pr_ren}
    P_r(x_1,x_2,\cdots, x_N,t) = \langle x_1, x_2, \dotsc, x_N|\hat\varrho_r(t) |x_1, x_2, \dotsc, x_N\rangle.
\end{equation} 
Therefore, it follows from Eqs.~\eqref{eq:rho_ren} and \eqref{eq:Pr_ren} that
\begin{equation}
\mspace{-8mu}
P_r(x_1,x_2,\cdots, x_N,t)=  e^{-r t} P(x_1,x_2,\cdots, x_N,t) + r\int_0^t d\tau\, e^{-r \tau} P(x_1,x_2,\cdots, x_N,\tau)\, , 
\label{eq:ren}
\end{equation}
where $P(x_1,x_2,\cdots, x_N,t)$ is the JPDF of the positions in the absence of resetting and is given by 
\begin{equation}
\mspace{-8mu}
    P(x_1,x_2,\cdots, x_N,t)= |\Psi(x_1,x_2,\cdots, x_N,t)|^2 = \langle x_1, x_2, \dotsc, x_N|\hat\varrho(t) |x_1, x_2, \dotsc, x_N\rangle.
    \label{eq:dm}
\end{equation}
In the limit $t\to\infty$, the first term in Eq.~\eqref{eq:ren} drops out and one arrives at a nonequilibrium steady state (NESS) given by 
\begin{equation}
P_r(x_1,x_2,\cdots, x_N)=   r\int_0^\infty d\tau\,e^{-r \tau} P(x_1,x_2,\cdots, x_N,\tau)\, , 
\label{eq:ren_NESS}
\end{equation}
where $P(x_1,x_2,\cdots, x_N,\tau)$ is given in Eq.~\eqref{eq:dm}.

Therefore, from \eref{eq:ren_NESS}, one sees that to obtain the JPDF of the positions of the particles in the NESS induced by stochastic   resetting, it is necessary to know the many-body time-dependent wavefunction
$\Psi(x_1,x_2,\ldots,x_N,t)$ without resetting, but  at all times $t$ (and not just at late times).

The average density profile of the gas (normalized to unity) in the presence of resetting is defined as
\begin{equation}
    \rho_r(x,t)  = \left \langle\frac{1}{N}
    \sum_{i=1}^N\delta\left(x-x_i\right)
    \right\rangle ,
    \label{eq:density-def}
\end{equation}
where the average $\langle \cdot\rangle$  is to be evaluated with respect to the JPDF $P_r(x_1, x_2, \dotsc, x_N, t)$ of the positions in the presence of resetting. Thus $\rho_r(x,t)\, dx$ gives the average fraction of particles of the gas at time $t$ in the interval $[x,x+dx]$. Assuming that the JPDF $P_r(x, x_2, \dotsc, x_N,t)$ is invariant under a permutation of the $x_i$'s, it follows that the average density $\rho_r(x,t)$ in Eq.~\eqref{eq:density-def} is given by the 
marginal one point function 
\begin{equation}
    \rho_r(x,t) = \int P_r(x, x_2, \dotsc, x_N,t)\, dx_2\dots dx_N.
    \label{eq:rhor-1}
\end{equation}
The renewal equation \eqref{eq:ren} leads to the renewal relation for the density profile 
\begin{equation}
\label{eq:den_ren}
    \rho_r(x,t) = e^{- r t} \, \rho(x,t) + r\int_{0}^{t} \, d\tau \, e^{-r \tau } \rho(x,\tau) \, , 
\end{equation}
where 
\begin{equation}
    \rho(x,t) = \int |\Psi(x,x_2,\cdots, x_N,t)|^2\, dx_2\dots dx_N,
\end{equation}
is the marginal PDF of the positions in the absence of resetting. 

Thus to evaluate even the macroscopic density profile $\rho_r(x,t)$, we need the knowledge of $\rho(x,t)$, which in turn, requires the knowledge of the full many-body wavefunction $\Psi(x,x_2,\cdots, x_N,t)$. However,  obtaining the many-body time-dependent wavefunction is, in general, extremely challenging  for any interacting quantum many-body system.  Nonetheless, in the large $N$ limit, the density profile $\rho(x,t)$ can be found for a quantum system of weakly interacting bosons in a condensate from the GPE starting from the Thomas-Fermi initial state [see Eq.~\eqref{eq:rscale}]. 

In the limit $t\to\infty$, the system reaches a nonequilibrium steady state, where the density profile can be obtained by taking limit $t\to\infty$ in Eq.~\eqref{eq:den_ren}, i.e., 
\begin{equation}
\label{eq:den_ren_ss}
    \rho_r(x) =  r\int_{0}^{\infty} \, d\tau \, e^{-r \tau } \rho(x,\tau) \, .
\end{equation}

\section{Full counting statistics in the presence of resetting}
\label{app:FCS}

Let us start from Eq.~\eqref{eq:Hkappa_0}, i.e., 
\begin{equation}
    H(\kappa) = \int_{1}^{\infty} d\lambda \, h(\lambda) \, \delta[\kappa - \kappa(\lambda)] \, ,
    \label{eq:Hkappa0}
\end{equation}
where, using Eq.~\eqref{eq:rscale} and Eq.~\eqref{eq:TF_density} in Eq.~\eqref{eq:NL}, $\kappa(\lambda)=N_L/N$ is given by  
\begin{equation}
    \kappa(\lambda) = \frac{3}{4 \, x_0 \, \lambda} \int_{-L}^{L} \, dx \, \left(1- \frac{x^2}{x_0^2 \lambda^2} \right) \Theta(x_0^2 \lambda^2 - x^2) \, .
\end{equation}
Performing the integral in the above equation, in terms of the dimensionless box-length $l=L/x_0$, we have
\begin{equation}
\label{eq:n_L(lam)}
    \kappa(\lambda) = \begin{cases}\displaystyle
        \frac{ l }{2 \lambda} \left(3 - \frac{l^2}{ \lambda^2} \right) &\quad \text{for} \quad \lambda > l \\
        \displaystyle
        1 &\quad \text{for} \quad  \lambda  \leq l\, .
    \end{cases}
\end{equation}

\subsubsection*{Case $l>1$:} For $l>1$, $H(\kappa)$ has a singular part coming from the plateau $\kappa=1$ in the integration range $1\le \lambda \le l$ in Eq.~\eqref{eq:Hkappa0},
\begin{equation}
    H_\text{sing}(\kappa)= \bigl[1-F(l)\bigr]\, \delta(\kappa-1),
\end{equation}
where $F(\lambda)$ is defined in Eq.~\eqref{eq:F}. The regular part of $H(\kappa)$ comes from the remaining range $l\le \lambda < \infty$ of the integral in ~\eqref{eq:Hkappa0},
\begin{equation}
   H_\text{reg}(\kappa)= \int _{l}^{\infty} \, d\lambda \, h(\lambda)\,  \delta{\left[ \kappa - \frac{l}{2 \lambda} \left( 3 - \frac{l^2}{ \lambda^2} \right) \right]} .
   \label{eq:Hreg}
\end{equation}
With $z=l/\lambda$, 
\begin{equation}
    \kappa = \frac{1}{2} (3 z- z^3),
    \label{eq:kap}
\end{equation}
is a depressed cubic equation in $z$. Since 
\begin{equation}
    \kappa'(\lambda) = \frac{3 z^2}{2l}(z^2-1) < 0, \quad\text{for}\quad z\equiv l/\lambda <1,  
\end{equation}
the map in Eq.~\eqref{eq:kap} is monotonic and the root is unique. Putting $z=2\sin\theta$ and using the identity $3\sin\theta - 4 \sin^3\theta=\sin3\theta$, Eq.~\eqref{eq:kap} reads $\sin 3\theta =\kappa$. Therefore, the unique root of Eq.~\eqref{eq:kap} is given by
\begin{equation}
    z_*(\kappa) = 2\sin\theta\quad\text{with}\quad \theta= \frac{1}{3}\arcsin(\kappa).
    \label{eq:lambdastarC}
\end{equation}
Consequently,
\begin{equation}
    \lambda_*(\kappa) = \frac{l}{2\sin\left[\frac{1}{3}\arcsin(\kappa)\right]},
    \label{eq:lambdastarC2}
\end{equation}
and the Jacobian from Eq.~\eqref{eq:kap},
\begin{equation}
    |\kappa'(\lambda_*)| =  \frac{3 z_*^2}{2l}(1-z_*^2)= \frac{6\sin^2\theta}{l\cos\theta}\, \sqrt{1-\kappa^2}\, , 
\end{equation}
where we have used the identity $\sqrt{1-\kappa^2}=\cos3\theta= \cos\theta (1-4\sin^2\theta)$. Therefore, performing the integral in Eq.~\eqref{eq:Hreg} yields
\begin{equation}
    H_\text{reg}(\kappa)= \frac{h(\lambda_*)}{|\kappa'(\lambda_*)|}, \quad 0<\kappa<1.
    \label{eq:HkappaC}
\end{equation}
Using Eq.~\eqref{eq:F}, it is easy to check the normalization $\int_0^1 H_\text{reg}(\kappa)\, d\kappa = \int_l^\infty h(\lambda)\, d\lambda = F(l)$. Therefore, $H(\kappa)=H_\text{sing}(\kappa) + H_\text{reg}(\kappa)$ satisfies the normalization $\int_0^1 H(\kappa)\, d\kappa =1$, as it must.

\subsubsection*{Case $l<1$:} For $l<1$, the plateau $\kappa(\lambda)=1$ in Eq.~\eqref{eq:n_L(lam)} does not contribute to the integral in Eq.~\eqref{eq:Hkappa0}. Therefore, there is no singular part in $H(\kappa)$. Moreover, the lower limit of integration for the regular part is $\lambda=1$ (instead of $l$) since $l<1$ is outside the support of $h(\lambda)$. Consequently, $\kappa$ has an upper support given by [see Eq.~\eqref{eq:n_L(lam)}]
\begin{equation}
 \kappa_{\max} = \kappa(1)= \frac{1}{2}(3l-l^3) <1.   
 \label{eq:kappamaxC}
\end{equation}
Nevertheless, the analysis of the regular part for the case $l>1$ still goes through for $\lambda<1$, except the normalization $\int_0^{\kappa_{\max}} H_\text{reg}(\kappa)\, d\kappa = \int_1^\infty h(\lambda)\, d\lambda = F(1) =1$.

\subsubsection*{Asymptotic behavior near the lower support.---} For $\kappa\to 0$, from Eq.~\eqref{eq:lambdastarC}, we have $\theta\simeq \kappa/3$, and consequently, from Eq.~\eqref{eq:lambdastarC2}, $\lambda_*(\kappa)\simeq 3l/(2\kappa)$. Therefore, using the large $\lambda$ asymptotic of $h(\lambda)$ from Eq.~\eqref{eq:h(lambda)_lim} in Eq.~\eqref{eq:HkappaC}, we get
\begin{equation}
    H(\kappa)\xrightarrow{\kappa\to 0} \frac{3r_0l}{2} \left(\frac{e}{6l}\right)^{r_0/2}\, \kappa^{r_0/2-2}\, \exp\left(-\frac{3r_0l}{2\kappa}\right)\, .
\end{equation}

\subsubsection*{Asymptotic behavior near the upper support.---} The $\kappa\to \kappa_{\max}$ limit needs to be evaluated separately for (a) $l>1$, (b) $l=1$, and (c) $l<1$. 

(a) For $l>1$, we have $\kappa_{\max}=1$. In this case, from Eq.~\eqref{eq:lambdastarC}, $\theta\to \pi/6$ as $\kappa\to 1$, and consequently, from Eq.~\eqref{eq:lambdastarC2}, $\lambda_*(1)=l$. Therefore, from Eq.~\eqref{eq:HkappaC}, we get 
\begin{equation}
    H_\text{reg}(\kappa) \xrightarrow{\kappa\to 1} \frac{l\, h(l)}{\sqrt{6}}\, \frac{1}{\sqrt{1-\kappa}}\, .
\end{equation}

(b) At $l=1$, for $\kappa\to 1$, we need to retain the next order term in $\lambda_*(\kappa)$ in Eqs.~\eqref{eq:lambdastarC}, i.e., $  \lambda_*(\kappa) \to   ( \, 1 + \sqrt{2/3}\, \sqrt{1-\kappa} \, )$. Now, using the expression of $h(\lambda)$ from Eq.~\eqref{eq:h_lam} in Eq.~\eqref{eq:HkappaC}, we get 
\begin{equation}
    H_\text{reg}(\kappa)\xrightarrow{\kappa\to 1} \frac{r_0}{3}\, \left(\frac{3}{2}\right)^{3/4}\, \frac{1}{(1-\kappa)^{3/4}}\, \approx \frac{(0.451801\dots) \, r_0}{(1-\kappa)^{3/4}}\, .
\end{equation}

(c) For $l<1$, $\kappa_{\max}$ is given by Eq.~\eqref{eq:kappamaxC}. Putting $\sin\theta = l/2$ in the identity, $3\sin\theta - 4 \sin^3\theta=\sin3\theta$, we get $\frac{1}{3}\arcsin(\kappa_{\max}) = \arcsin(l/2)$. Therefore, from Eq.~\eqref{eq:lambdastarC}, $\theta\to \arcsin(l/2)$ as $\kappa\to \kappa_{\max}$, and consequently, from Eq.~\eqref{eq:lambdastarC2},  $\lambda_*(\kappa_{\max})=1$, which is expected as that was how $\kappa_{\max}$ in Eq.~\eqref{eq:kappamaxC} was obtained. However, since $h(\lambda)\simeq r_0/\sqrt{\lambda-1}$ diverges as $\lambda\to 1$ [see Eq.~\eqref{eq:h_lam}], we need to keep the next order term in $\lambda_*(\kappa)$ as $\kappa\to \kappa_{\max}$. This can be easily found as $\lambda(\kappa)-1 \simeq 2(\kappa_{\max}-\kappa)/[3 l (1-l^2)]$, which yields
\begin{equation}
H(\kappa) \xrightarrow{\kappa\to\kappa_{\max}} r_0 \sqrt{\frac{2}{3 l \left(1-l^{2}\right)}} \,
\frac{1}{\sqrt{\kappa_{\max}-\kappa}}\, .
\label{eq:edge-lsmall}
\end{equation}

\section*{References}

\bibliographystyle{iopart-num-custom}
\bibliography{ref}

\end{document}